\documentclass[sigplan,screen]{acmart}
\AtBeginDocument{%
  }
\usepackage{pifont}
\usepackage{multirow}
\usepackage{makecell}
\usepackage{subcaption}

\def\cu#1{{\bfseries #1}}
\def\xi#1{\textit{#1}}

\setcopyright{acmlicensed}
\copyrightyear{2018}
\acmYear{2018}
\acmDOI{XXXXXXX.XXXXXXX}
\acmConference[Conference acronym 'XX]{Make sure to enter the correct
  conference title from your rights confirmation email}{June 03--05,
  2018}{Woodstock, NY}

\acmISBN{978-1-4503-XXXX-X/2018/06}

\def\cu#1{{\bfseries #1}}
\def\xi#1{\textit{#1}}

\usepackage{hyphenat}

\begin{document}

\title{ARSP: Automated Repair of Verilog Designs via Semantic Partitioning}


\author{Bingkun Yao}
\affiliation{%
  \institution{City University of Hong Kong}
  \city{Hong Kong}
  \country{China}}
\email{bingkun.yao@cityu.edu.hk}

\author{Ning Wang}
\affiliation{%
  \institution{City University of Hong Kong}
  \city{Hong Kong}
  \country{China}
}\email{nwang227-c@my.cityu.edu.hk}

\author{Xiangfeng Liu}
\affiliation{%
 \institution{City University of Hong Kong}
 \city{Hong Kong}
 \country{China}}
\email{xiangfengliu.cn@gmail.com}

\author{Yuxin Du}
\affiliation{%
  \institution{City University of Hong Kong}
  \city{Hong Kong}
  \country{China}}
  \email{yuxindu8-c@my.cityu.edu.hk}

\author{Yuchen Hu}
\affiliation{%
  \institution{Southeast University}
  \city{Nanjing}
  \country{China}}
\email{3042817243@qq.com}

\author{Hong Gao}
\affiliation{%
  \institution{Zhejiang Normal University}
  \city{Jinhua}
  \country{China}}
\email{honggao@zjnu.edu.cn}

\author{Zhe Jiang}
\affiliation{%
  \institution{Southeast University}
  \city{Nanjing}
  \country{China}}
\email{101013615@seu.edu.cn}

\author{Nan Guan}
\affiliation{%
  \institution{City University of Hong Kong}
  \city{Hong Kong}
  \country{China}}
\email{nanguan@cityu.edu.hk}

\renewcommand{\shortauthors}{Trovato et al.}

\begin{abstract}
Debugging functional Verilog bugs consumes a significant portion of front-end design time. While Large Language Models (LLMs) have demonstrated great potential in mitigating this effort, existing LLM-based automated debugging methods underperform on industrial-scale modules. A major reason for this is bug signal dilution in long contexts, where a few bug-relevant tokens are overwhelmed by hundreds of unrelated lines, diffusing the model’s attention.
To address this issue, we introduce ARSP, a two‑stage system that mitigates dilution via semantics‑guided fragmentation. A Partition LLM splits a module into semantically tight fragments; a Repair LLM patches each fragment; edits are merged without altering unrelated logic. A synthetic data framework generates fragment‑level training pairs spanning bug types, design styles, and scales to supervise both models. Experiments show that ARSP achieves 77.92\% pass@1 and 83.88\% pass@5, outperforming mainstream commercial LLMs including Claude‑3.7 and SOTA automated Verilog debugging tools Strider and MEIC. Also, semantic partitioning improves pass@1 by 11.6\% and pass@5 by 10.2\% over whole‑module debugging, validating the effectiveness of fragment‑level scope reduction in LLM-based Verilog debugging.
\end{abstract}

\keywords{Debugging, Verification, Verilog, Large language model (LLM)}


\received{20 February 2007}
\received[revised]{12 March 2009}
\received[accepted]{5 June 2009}

\maketitle

\section{Introduction}
In modern ASIC/FPGA development, an undetected functional bug in Verilog design can lead to Engineering Change Orders (ECOs) and chip re-spins, resulting in substantial economic losses. Debugging functional bugs is time-consuming--once a bug is exposed, engineers need to manually analyze the root cause within hundreds of lines with intertwined sequential and combinational logic, then carefully repair the code according to the design specifications. Research \cite{debuggingtime} indicates that functional bug debugging can consume up to two-thirds of chip front-end design time. Thus, efficient automated Verilog debugging methods are of significant value. Given the exceptional code comprehension capabilities demonstrated by Large Language Models (LLMs) in recent years\cite{llmcoding,llmcoding2}, leveraging LLMs to assist Verilog debugging has emerged as a natural direction.

Recent work has applied LLMs for automated Verilog debugging \cite{TCS,TIFS24,veriassist,sp23,todaes25,rtlfixer,uvllm,aivril,veridebug,meic,hdldebugger}, yet the fix quality on industrial-scale modules remains unsatisfactory. A crucial obstacle is what we call \xi{bug signal dilution}: in long hardware modules, only a small fraction of tokens is actually bug-related, making it difficult for the model to focus on the true bug location. Sec. \ref{sec:bugsignal} details this problem.

We empirically find that Verilog bugs are highly localized: in our manual study across representative debugging benchmarks and an industrial dataset, most bugs reside entirely within a single \xi{semantically tight fragment}—a small, cohesive region whose declarations, signals, and control predicates implement one micro-function (see Sec. \ref{sec:buglocal}). This strong locality motivates our fragmentation-based strategy: decompose a module into semantically tight fragments, repair fragments individually, and merge edits to mitigate bug signal dilution.


In this paper, we present ARSP (Automated Repair via Semantic Partitioning), a system that couples semantics-guided fragmentation with staged LLM debugging. ARSP mainly uses two specialized LLMs that we train for their tasks: (1) a \cu{Partition LLM} that splits a module into semantically tight fragments, and (2) a \cu{Repair LLM} that produces a patch for each fragment. All edits are merged to form the repaired module. 
To address the lack of training data for module fragmentation and repair, we build a data synthesis framework that generates synthetic fragment‑level training examples covering common bug types, varied types of hardware designs, and different code scales. These synthetic pairs directly support our training. We focus on functional bugs, which are bugs that pass syntax checking and logic synthesis but fail to meet the intended design specifications.

To the best of our knowledge, ARSP is the first Verilog debugging approach that mitigates bug signal dilution caused by long contexts by using semantics-guided code partitioning. It requires no testbench, SVA, reference model, or other verification components, thus reducing deployment overhead. ARSP achieves pass@1 and pass@5 repair success rates of 77.92\% and 83.88\% respectively on our test dataset consisting of industrial-scale Verilog modules, outperforming leading commercial LLMs including Claude-3.7 and Deepseek and SOTA automated Verilog debugging methods. Ablation studies show that semantic partitioning improves pass@1 by 11.6\% over whole-module debugging, underscoring the effectiveness of fragment-level scope reduction.

\section{Motivation}
\subsection{Bug Signal Dilution in Long Contexts}\label{sec:bugsignal}
Existing work has attempted to leverage LLMs for automated Verilog debugging \cite{TCS,TIFS24,veriassist,sp23,todaes25,rtlfixer,uvllm,aivril,veridebug,meic,hdldebugger}. However, these approaches still fall short of producing acceptable code fixes for industrial-scale Verilog module designs. While the scarcity of high‑quality training data is a widely acknowledged cause of this issue, we identify a single dominant cause that would persist even under hypothetical data abundance: Bug signal dilution--a substantial sparsity of defect‑relevant tokens within long contexts of Verilog design.
\begin{figure}[!htb]
  \centering
\includegraphics[width=\linewidth]{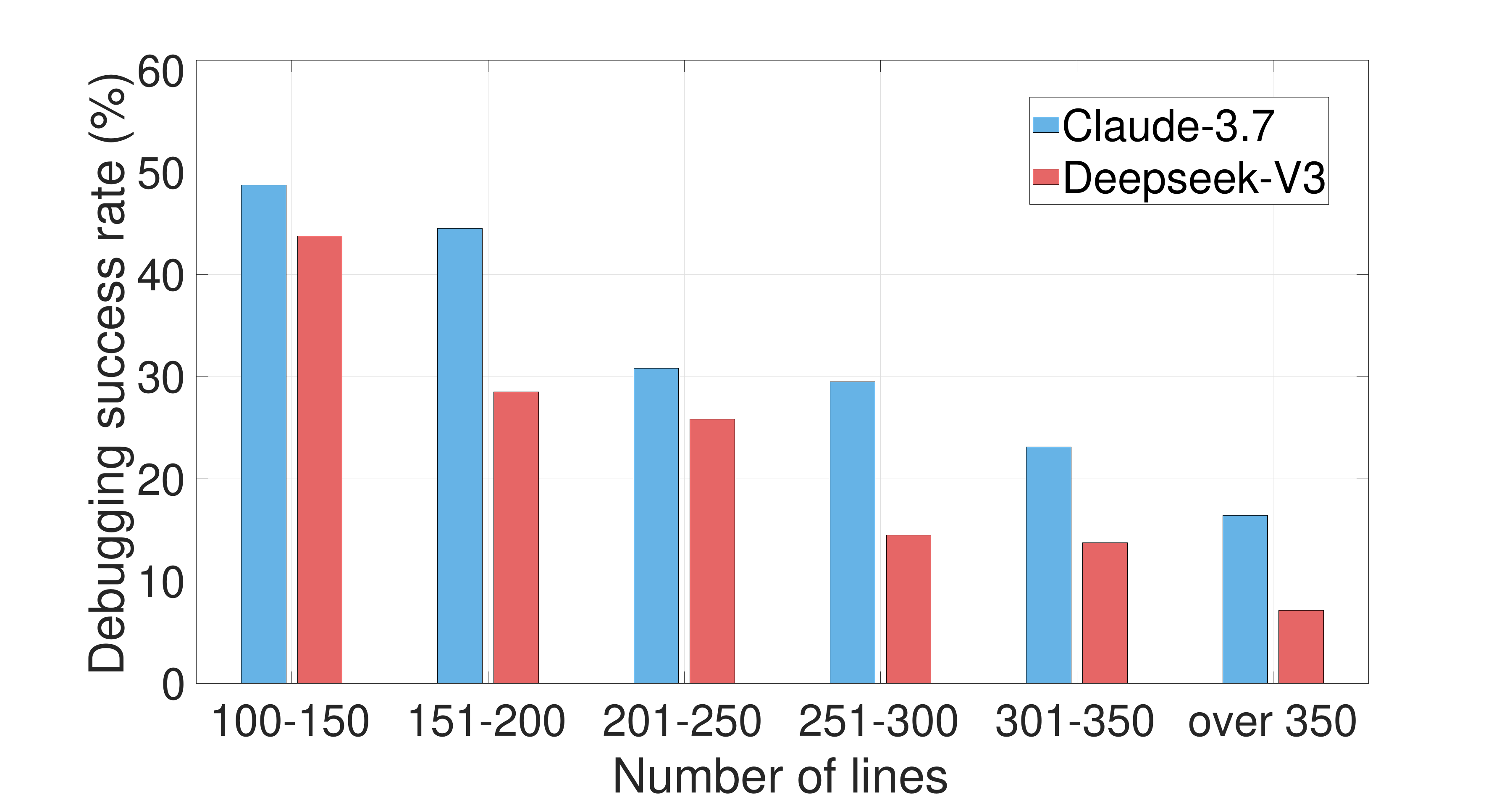}
  \caption{On our test dataset constructed from GenBen\cite{genben}, which is a benchmark on LLM-based Verilog coding tasks, the debugging success rate decreases as the code scale increases.}\label{fig:successratevsscale}
\end{figure}

Industrial Verilog modules typically span hundreds of lines of intertwined sequential and combinational logic, yet actual bugs often involve only a few lines. Our analysis of a proprietary industrial dataset containing 124 historical bug repair patches reveals that 115 patches (92.7\%) modify no more than 3 lines, with 104 patches (83.9\%) involving single-line modifications. As the line count increases, the proportion of bug-related tokens in the overall context rapidly decreases, causing model's attention to be diluted across large swaths of semantically irrelevant code, which degrades the debugging success rate (Fig. \ref{fig:successratevsscale}). Addressing bug signal dilution therefore motivates the design principles behind our approach.

\subsection{Locality of Verilog Bugs}\label{sec:buglocal}

\begin{figure}[t]
  \centering
\includegraphics[width=\linewidth]{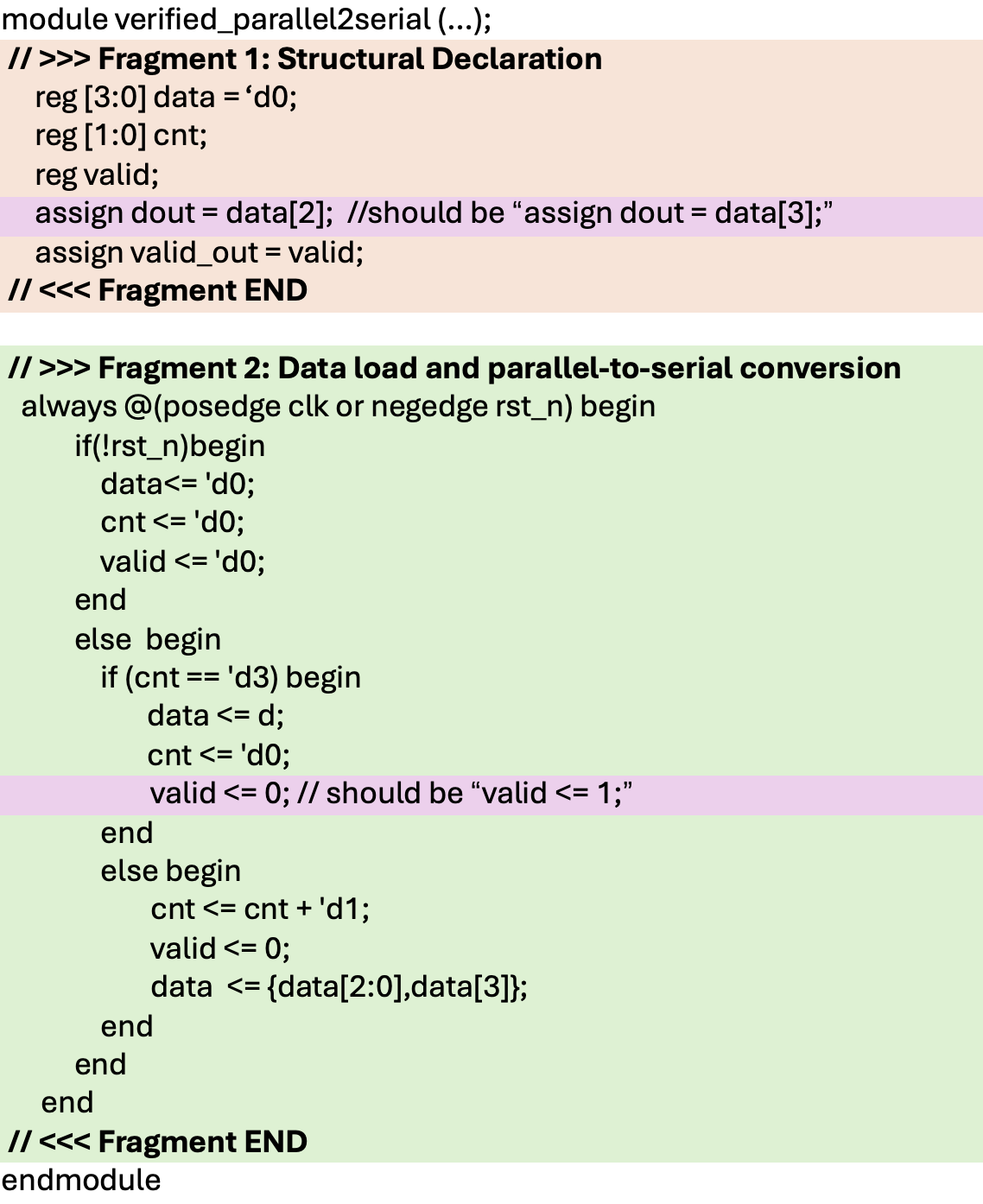}
  \caption{Examples of bugs in semantic tight fragments. The first bug is a datapath error located in Fragment 1, while the second bug is a control logic error located in Fragment 2.}\label{fig:fragment}
\end{figure}

Empirically, we observe that most Verilog bugs arise within a small, semantically tight fragment. We define a semantically tight fragment as \xi{a localized region whose internal elements (declarations, signals, control predicates) collaborate to implement one coherent micro-function.} Fig. \ref{fig:fragment} shows an example of bugs within such fragments from the test dataset of MEIC \cite{meic}, the SOTA LLM-based Verilog debugging method. The module implements parallel-to-serial conversion, where every four input bits are converted to a serial one-bit output. It can be divided into two semantic fragments. Fragment 1 declares the state and bindings. It contains a datapath bug: the output signal dout is incorrectly routed to data[2]. Fragment 2 implements data loading and parallel-to-serial conversion, exhibiting a control logic error where the valid signal fails to assert after loading. Both bugs reside entirely inside a semantic tight fragment. 

To verify the locality of bugs, we conduct a manual analysis of 145 bugs across 42 buggy modules from benchmarks of representative research works on automated Verilog debugging including Cirfix\cite{cirfix}, RTL-Repair\cite{rtlrepair}, Strider\cite{strider} and MEIC\cite{meic} as well as a proprietary industrial dataset from our partner company. Two hardware engineers with over five years of experience independently identify the minimal semantic fragment for each bug. To avoid circular reasoning, we first ask the engineers to partition each buggy module into semantic tight fragments without knowing the locations of the bugs. Then, for each bug, we check whether it falls within a single fragment. The Cohen's Kappa coefficient between the two engineers is 0.7434, indicating substantial agreement above the threshold of 0.6 \cite{cohenkappa}. Disagreements were resolved through discussion. The result shows that 142 out of 145 bugs (97.93\%) are entirely contained within a single semantic tight fragment. These fragments are generally small in scale: the median line count is 34, average line count is 28.3, and the 90th percentile is 51 lines. The average proportion of a single fragment to the entire module is 14.6\%. This provides strong empirical evidence of bug locality. For transparency and to facilitate future research, we release some examples of semantic tight fragments and the bug–fragment mappings at \url{https://anonymous.4open.science/r/ARSP-B10D.} 

Given this strong locality, we need not rely on an LLM to stay focused inside a large context. Instead, we can split a module into multiple smaller semantic fragments, repair each fragment separately, and then merge the edits back, keeping the module semantically equivalent beyond the repaired region. 
During this process, LLMs do not require every fragment to be a complete standalone module. As long as a fragment forms a semantically independent block, the LLM can repair it. This fragment-based process directly targets bug signal dilution by shrinking the context from a large, diffuse module to small, high‑signal fragment scopes.
\section{The Design of ARSP}
\subsection{ARSP Overview}
\begin{figure*}[t]
  \centering
  \includegraphics[width=\linewidth]{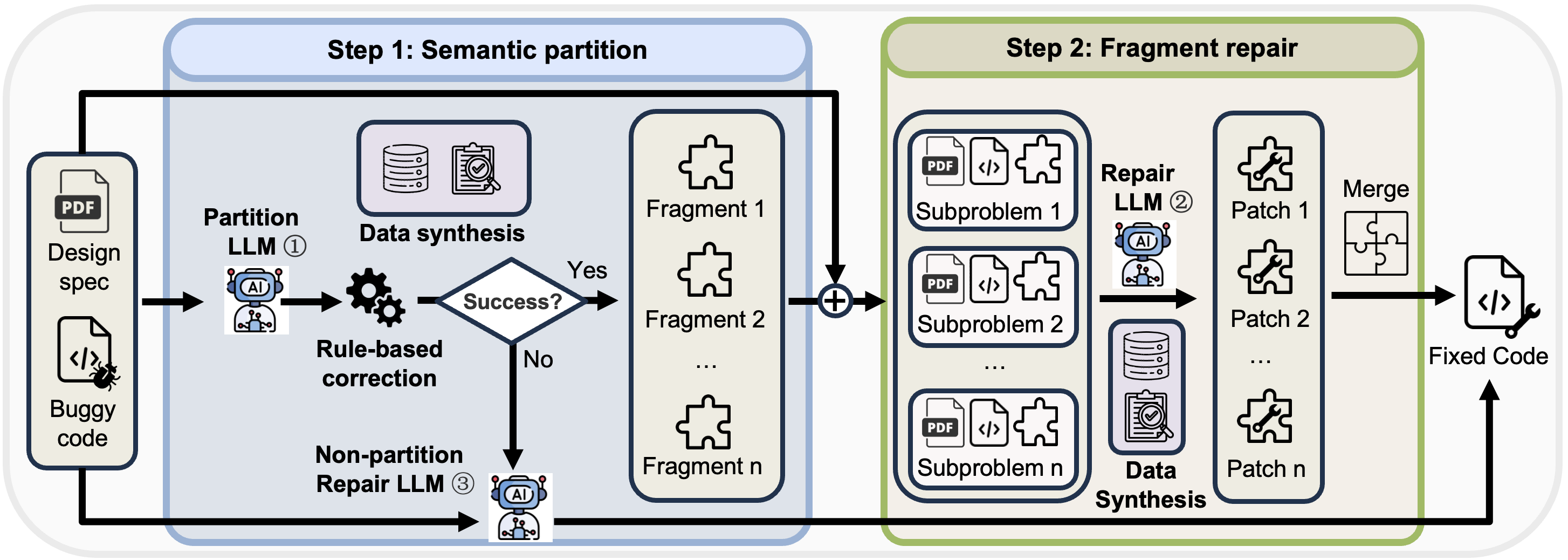}
  \caption{The workflow of ARSP, including two steps: (1) Semantic partition to split the code into several semantic tight fragments and (2) Fragment repair to generate patch for each fragment. We also propose a data synthesis framework to generate the training data for LLMs used in the two steps.}\label{fig:workflow}
\end{figure*}

Fig. \ref{fig:workflow} shows the workflow of ARSP. It takes the design specification and the buggy code as input and output the fixed code. ARSP consists of two steps: Semantic partition and fragment repair.

\cu{(1) Semantic partition:} This step splits the buggy code into several semantic tight fragments. Since we require each fragment to be a localized region whose internal elements collaborate to implement a coherent micro-function, methods based on fixed rules are not suitable here. Thus, we employ an LLM, namely \cu{Partition LLM} to partition the buggy code according to the design specification. Due to the stochastic nature of LLM outputs, we apply a rule-based method to process the output of Partition LLM to ensure that the concatenation of all generated fragments exactly matches the original buggy code. If partition fails, we use an LLM called \cu{Non-Partition Repair LLM} to fix the code, which directly repairs the complete buggy code without partition. 

\cu{(2) Fragment repair:} An LLM called \cu{Repair LLM} repairs each fragment. To allow the model to focus on one designated fragment while still retaining global semantic context, we provide (i) the design specification, (ii) the complete buggy code, and (iii) the target fragment to the Repair LLM, and instruct it to concentrate on the designated fragment and output only the repaired version of that fragment. Then, all repaired fragments are concatenated to form the fixed code. 

Moreover, existing open-source datasets on Verilog debugging are very scarce and are mainly confined to short code snippets under 100 lines, which do not match the complexity of industrial code. Thus, we propose a data synthesis pipeline to produce high-quality data with industrial-level complexity tailored to the Verilog debugging task to train the model. 


\subsection{Step 1: Semantic Partition}

\subsubsection{Training Process of Partition LLM} 
Due to the prohibitive computational cost and requirements of massive high-quality data, it is infeasible for us to train Partition LLM from the scratch. Fortunately, some open-source base LLMs specialized for coding tasks are now available. They have been trained on massive corpora related to coding tasks and thus possess broad foundation knowledge of such tasks. However, since this training is unsupervised, they can not answer specific user questions\cite{llmsurvey,llmsurvey2}. Therefore, we perform supervised training to equip the model with the capability of code partition. We choose Deepseek-Coder-6.7B \cite{deepseekcoder} as the base LLM for its strong programming knowledge; more importantly, its release date (October 2023) precedes the release of our testsets (Novemver 2023 and May 2025), thereby avoiding potential data contamination. 

In our supervised training, each training sample is an input-output pair (\xi{q$_s$}, \xi{a$_s$}), where \xi{q$_s$} is the problem of semantic partition, containing the complete buggy code as well as its design description. \xi{a$_s$} is the desired answer, namely all semantic tight fragments. To reduce training cost, we use LoRA \cite{lora} as the training algorithm, which reduces the computational and memory overhead by updating only a small subset of the model parameters.

\subsubsection{Training Data of the Partition LLM} We need (\xi{q$_s$}, \xi{a$_s$}) pairs to train the Partition LLM. Dividing the Verilog code into semantic tight fragments is a new, highly specific task, which has not been addressed in prior work. Consequently, no publicly dataset is available, and manually creating large-scale training data is prohibitively expensive. Thus, we synthesis the (\xi{q$_s$}, \xi{a$_s$}) pairs for semantic partition using a powerful commercial LLM. The data synthesis consists of two steps: (1) Code partitioning: we use the input of semantic partition problem \xi{q$_s$} as the prompt to the LLM to obtain \xi{a$_s$}, the resulting code fragments. (2) Data filtering: we remove the (\xi{q$_s$}, \xi{a$_s$}) pairs with incorrect or low-quality \xi{a$_s$} to ensure the quality of the training data.

\cu{(1) Code partitioning:} We use Claude-3.7 \cite{claudeapi}, a mainstream commercial LLM with strong performance on coding tasks. We formulate the prompt \xi{q$_s$} as:
\begin{equation}
    \xi{t} \circ \xi{g} \circ \xi{o} \circ \xi{i},
\end{equation}
where $\circ$ is the concatenate operator, \xi{t} represents the specification of semantic partition problem that defines the objective; \xi{g} is a guidance detailing how to partition the code; \xi{o} specifies the output format, and \xi{i} denotes the problem input including the complete buggy code and the design specification. 

\begin{figure*}[t]
  \centering
\includegraphics[width=\linewidth]{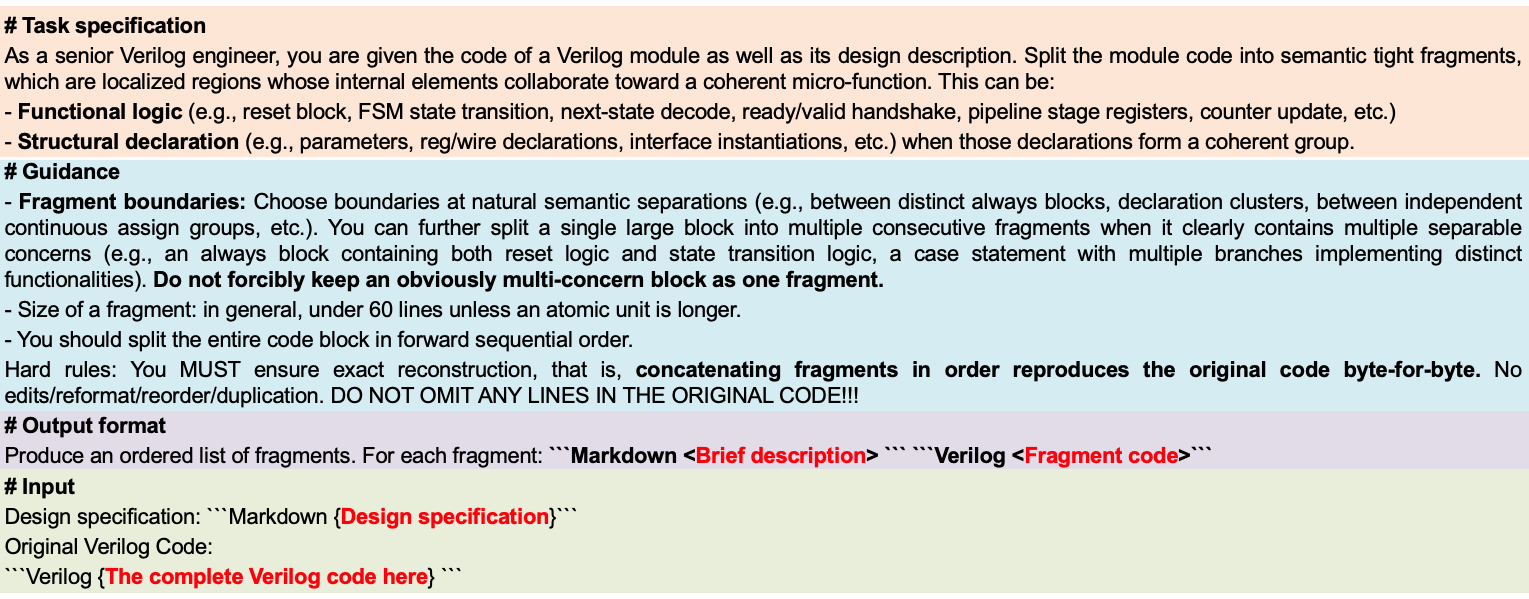}
  \caption{The prompt for Verilog semantic partition, including the task specification, guidance and the input/output format.}\label{fig:prompt_partition}
\end{figure*}

Fig. \ref{fig:prompt_partition} shows \xi{q$_a$} in detail. In the task specification, we guide the model to consider two types of semantic tight fragments: one type implements specific micro-functions, such as reset clocks, FSM state transition and next-state decode. The other is declarative, consisting of structural declarations such as parameters, wire/register definitions, and interface instantiations. This binary categorization comprehensively captures all semantic tight fragments. 

In the task guidance, we instruct the model to choose fragment boundaries at natural semantic separations, such as distinct always blocks. For large blocks that implement multiple functions, the model may split them into multiple fragments according to the semantics of those functions. For example, an always block containing both reset logic and state transition logic can be split into two fragments corresponding to these two functions. Furthermore, we require that concatenating the fragments in their original order yields code that is exactly identical to the original. To preserve the original code’s functionality, we require that concatenating the fragments in their original order yields code that is exactly identical to the original. 

Finally, to facilitate verification of the partition results, we require each fragment's code to be enclosed between "\verb|```|Verilog" and "\verb|```|", and a brief description on the micro-function of each fragment to be included in the output.

\cu{(2) Data filtering.} Basically, to preserves the original code's functional behavior, we require that combining the fragments in \xi{a$_s$} by their original order exactly reproduces the original buggy code in \xi{q$_s$}; any (\xi{q$_s$}, \xi{a$_s$}) pair failing this check is discarded. 

Then, to maintain a practical partition granularity, we exclude (\xi{q$_s$}, \xi{a$_s$}) pairs whose fragment count exceeds 15 or that contain any fragment longer than 80 lines, preventing both excessive segmentation and overly large fragments. 

Finally, we need to remove (\xi{q$_s$}, \xi{a$_s$}) pairs with fragments not conforming to our definition of a semantic tight fragment, i.e., its internal elements implement a coherent micro-function. This notion is intrinsically qualitative and relies on expert judgment, making it difficult to verify using methods based on fixed rules. Inspired by prior research on LLM-as-a-judge \cite{llmasjudge1,llmasjudge2,llmasjudge3}, we leverage the strong language understanding capabilities of LLMs to perform this qualitative judgment. 
Specifically, we employ Claude-3.7 as the judge. We provide the complete buggy code, the design description, and the generated fragments with their brief descriptions, and ask the model whether all fragments satisfy the definition of semantic tight fragment. We require the model to provide an explicit "Yes" or "No" judgment at the end of its answer. To ensure the robustness of the judgment, we employ a voting mechanism: each (\xi{q$_s$}, \xi{a$_s$}) pair is evaluated 5 times, and if at least 3 of the judgments are ``Yes", the pair is regarded as correct. 

We synthesize a total of 24202 (\xi{q$_s$}, \xi{a$_s$}) pairs. On average, each buggy code sample is partitioned into 6.34 semantic tight fragments. Each fragment has an average length of 32.16 lines, accounting for 15.72\% of its complete buggy code.


\subsubsection{Rule-based Correction on the Partition Result}\label{sec:rulebasedcorrection}
Due to the inherent uncertainty of LLM outputs, the fragments produced by the Partition LLM may fail to reconstruct the original buggy code exactly. We examine 80 randomly sampled outputs whose fragments could not be concatenated into the original buggy code. 
We observe that some lines in certain fragments differ from their corresponding lines in the original buggy code, so the recomposed code of these fragments may no longer preserve the original semantics. 

\begin{figure}[t]
  \centering
  \includegraphics[width=\linewidth]{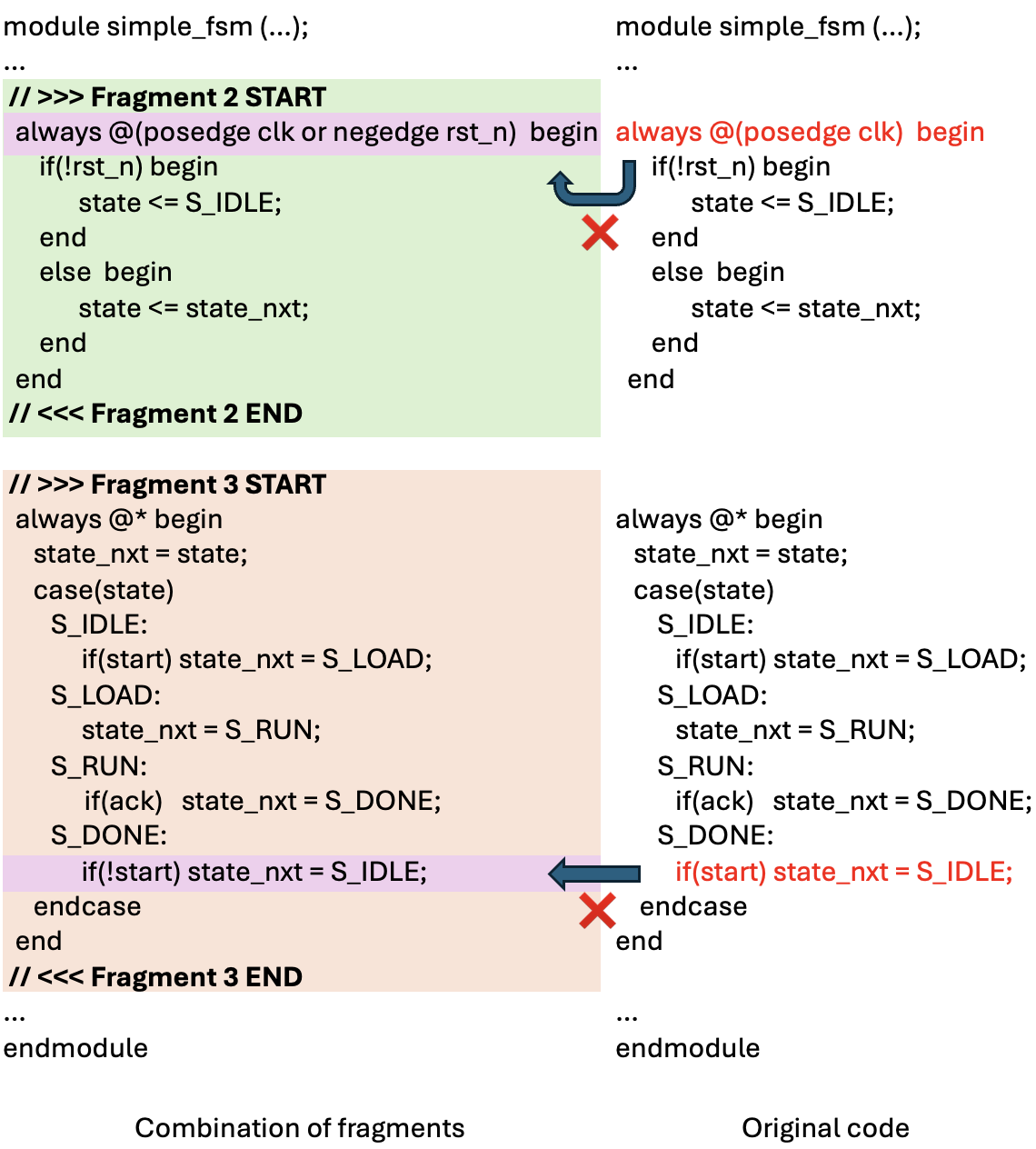}
  \caption{For some fragments output by the Partition LLM, certain lines are modified compared to their corresponding lines in the original code. Consequently, the recomposed code of these fragments may no longer preserve the original semantics.}\label{fig:fragmenterror}
\end{figure}

Fig. \ref{fig:fragmenterror} shows such a case: after partitioning, the sensitivity list of the always block in Fragment 2 is modified, in which the added rst\_n signal converts the reset operation from synchronous to asynchronous. In Fragment 3, the transition condition for S\_DONE changed from if(start) to if(!start), changing the module's handshake protocol. These alterations may stem from LLMs' hallucination in code generation, where the model over-confidently generates code that appears to be reasonable but diverges from user's requirements \cite{hallucination}. Given that subsequent debugging is performed on individual fragments, such deviations from the original code may misdirect the debugging process by masking the original defect or introducing false anomalies.

Fortunately, we find that for 72 of the 80 cases where the assembled fragments differ from the original code, the combination of all fragments has the same number of lines and identical structure as the original code. As shown by the example in Fig. \ref{fig:fragmenterror}, despite having different lines, the combination of fragments and original code maintain the same module structure. They have identical always block positions and equivalent case statement organization across corresponding lines. This alignment in code structure enables us to establish a direct line-by-line correspondence between the recomposed and original code. Therefore, we derive the fragment boundaries directly from their line spans. That is, if Fragment \xi{n} produced by the Partition LLM occupies lines \xi{n$_s$} through \xi{n$_e$} in the recomposed code, the corrected Fragment \xi{n} is also from line \xi{n$_s$} to \xi{n$_e$} in the original buggy code. Through this method, we can correct 90\% of the Partition LLM's outputs where the assembled fragments differ from the original code.

Through the approach above, if the combination of fragments output by the Partition LLM has the same number of lines as the original buggy code, the semantic partition is successful, and we proceed to repair each fragment individually. Otherwise, we employ the Non-Partition Repair LLM to directly fix the complete buggy code.

\subsubsection{Training of Non-Partition Repair LLM} We employ this LLM to repair the complete code in a single pass when partitioning fails. Similar to the Partition LLM, we use Deepseek-6.7B-Base\cite{deepseekcoder} as the base model and perform supervised training with LoRA\cite{lora} algorithm to equip it with debugging capabilities. Each training data sample can be modeled as (\xi{q$_n$}, \xi{a$_n$}) pairs, where \xi{q$_n$} is the input to the non-Partition LLM, including a prompt instructing the model to fix errors in the code, design descriptions, and the complete buggy code. \xi{a$_n$} denotes the expected output, namely the corrected complete code. 

\begin{figure}[h]
  \centering
\includegraphics[width=\linewidth]{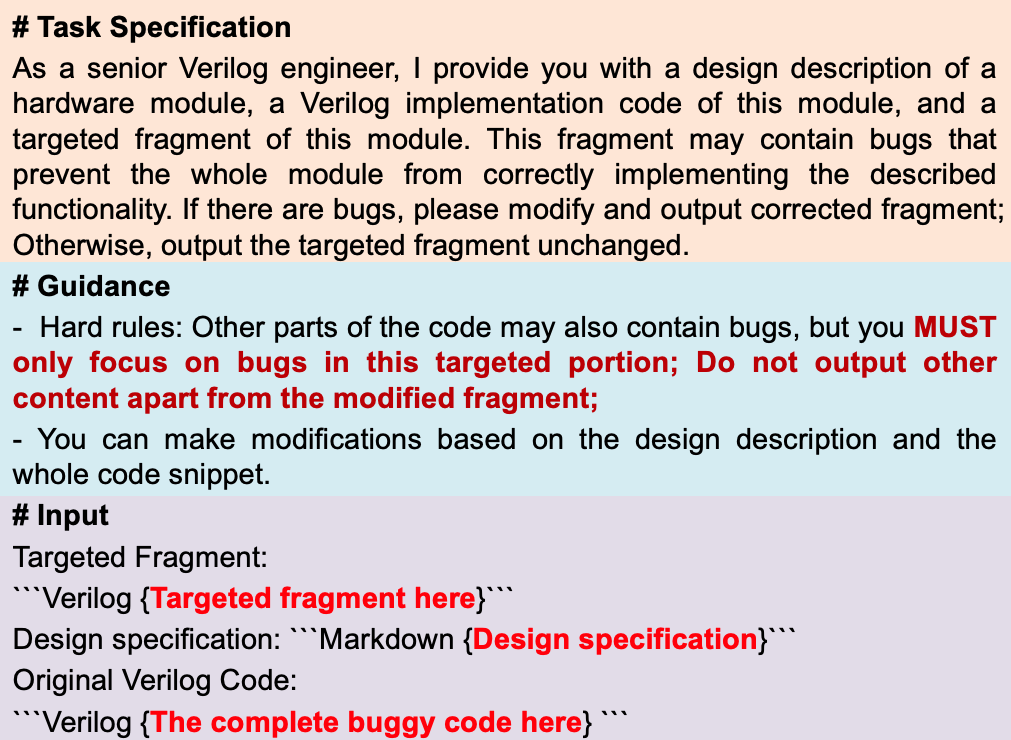}
  \caption{The prompt \xi{q$_r$} for Repair LLM.}\label{fig:prompt-debug}
\end{figure}

\subsection{Step 2: Fragment Repair}
\subsubsection{Training Process of the Repair LLM} 
After we obtain the semantic fragments of the complete code, we use the Repair LLM to fix each fragment, and then combine them to form the repaired complete code. Similar to the Partition LLM, we select Deepseek-Coder-6.7B as the base model and employ the LoRA for supervised training to obtain the Repair LLM. Each training data sample is a pair (\xi{q$_r$}, \xi{a$_r$}), where \xi{q$_r$} is the input to the Repair LLM and \xi{a$_r$} is the expected output.

The format of \xi{q$_r$} is shown in Fig. \ref{fig:prompt-debug}, which contains an instruction that require the model to fix a specified fragment. Additionally, to ensure that the model does not lose global information of the complete code when processing an individual fragment, we also include the complete buggy code and design description in \xi{q$_r$}. Meanwhile, we explicitly instruct the model to focus only on the given fragment and constrain the model's output \xi{a$_r$} to contain only the repaired fragment. In this way, we can mitigate the dilution of the bug signal in long contexts without discarding global contexts. 

\subsubsection{Training Data Synthesis} Regarding the input \xi{q$_r$}, we have already synthesized the training data for the Partition LLM, which consists of aligned design descriptions, complete buggy code and the fragments partitioned from the buggy code. As shown in Fig. \ref{fig:prompt-debug}, these components directly correspond to the required elements of \xi{q$_r$}.

Next, we discuss the expected output \xi{a$_r$}, which is the repaired fragment. Since the buggy code is synthesized by introducing bugs into the correct code, as will be discussed in Sec. \ref{sec:datasynthesis}, here we also have access to the correct code. Since we expect the correct code to be composed of repaired fragments of the buggy code, we need to identify the corresponding parts in the correct code for all fragments of the buggy code. These parts serve as the expected output \xi{a$_r$}. This correspondence is a semantic relationship, which is difficult for rule-based methods to capture. 

\begin{figure}[t]
  \centering
\includegraphics[width=\linewidth]{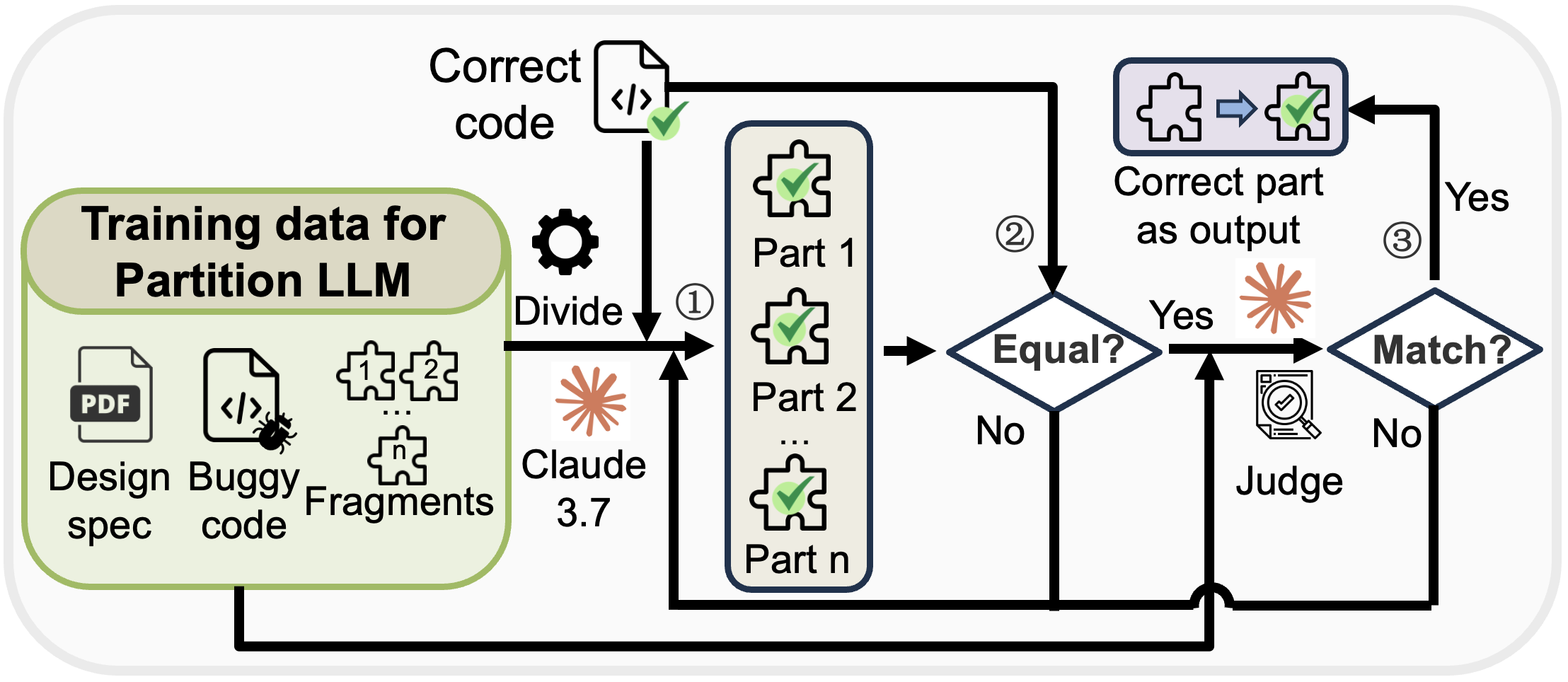}
\caption{Finding the corresponding parts in the correct code for each fragment in the buggy code. \textcircled{1}: Generate each parts of the correct code. \textcircled{2}: Check whether the combination of all parts equals the correct code. \textcircled{3}: Check the semantic correspondence between parts of the correct code and fragments of the buggy code.}\label{fig:generatecorrectfragment}
\end{figure}

Given the strong code understanding capabilities of LLMs, we use a powerful commercial LLM Claude-3.7 to identify the corresponding parts in the correct code for each fragment in the buggy code. This process is shown in Fig. \ref{fig:generatecorrectfragment} and consists of three steps: 

\cu{(1)} We provide the buggy code, each fragment of the buggy code and the correct code to Claude-3.7, and instruct it to divide the correct code into multiple parts. We require that these parts, in sequential order, be in a one-to-one semantic correspondence with the fragments of the buggy code. Moreover, when combined, these parts must be exactly equivalent to the original correct code.

\cu{(2)} Check whether the combination of all parts is exactly identical to the original correct code.

\cu{(3)} Use Claude-3.7 as a judge LLM, asking it whether each part of the correct code is in one-to-one semantic correspondence with each fragment of the buggy code. If the answer is "Yes", we consider all the parts of correct code to be correct and take them as the expected output \xi{a$_r$}. 

If the parts generated in Step (1) do not pass the checks in Step (2) and (3), we return to Step (1) to regenerate them. We limit the number of iterations to at most 5.

\subsection{Training Data Synthesis}\label{sec:datasynthesis}
To support our LLM training, we require aligned datasets consisting of the design description, the buggy code, and the corresponding corrected code. However, due to the proprietary of hardware design and the coarse granularity of contributions in open-source repositories, such aligned datasets are nearly unavailable in the current open-source community. Moreover, most open-source Verilog modules have fewer than 100 lines. This is smaller than typical industrial modules, which usually span several hundred lines. 

\begin{figure}[t]
  \centering
\includegraphics[width=\linewidth]{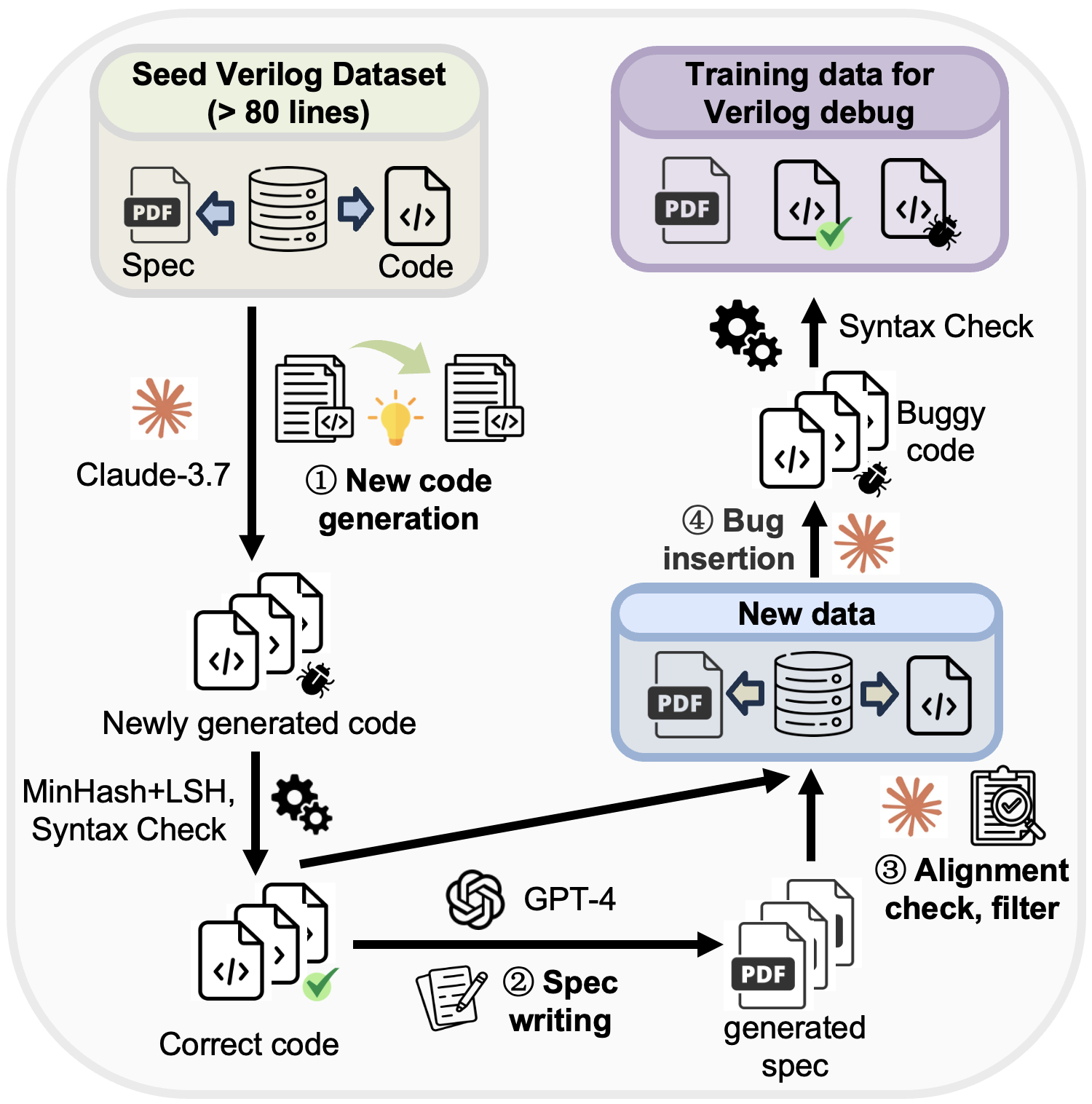}
\caption{The pipeline for synthesizing the training data on Verilog debugging.}\label{fig:generatedata}
\end{figure}

To address issues above, we propose a pipeline based on commercial LLMs to synthesis aligned training data for Verilog debugging. The generated designs have complexity comparable to industrial modules. The general idea is to generate new Verilog modules together with their design specifications based on a seed dataset, then create buggy code by injecting bugs into these modules. Fig. \ref{fig:generatedata} shows the entire pipeline. We detail each step below:

\cu{(1) Seed dataset.} Training datasets applied by existing studies on LLM-based Verilog generation generally contain aligned design specifications and code. Therefore, we choose datasets from several representative works in this field as the seed including Veri-Gen\cite{verigen}, VeriSeek\cite{veriseek}, PyraNet\cite{pyranet} and HaVen\cite{haven}. We apply MinHash\cite{minhash} and LSH\cite{lsh} algorithms to remove duplicate modules and filter those modules with syntax errors. Regarding the code scale, modules containing fewer than 100 lines lack sufficient complexity. Modules over 400 lines are very likely to exceed the context window of our base model (16384 for Deepseek-Coder-6.7B), causing truncation during training, making the design specification become misaligned with the partial code provided. Thus, we retain modules with 100–400 lines of code. The resulting seed dataset contains 9685 modules.

\cu{(2) New code generation.} We adopt OSS-Instruct\cite{OSS-Instruct}, a prompting strategy that stimulates the model to generate diverse training data, to produce new Verilog modules. For each item in the seed dataset, we input its design specification and code into Claude-3.7 and instruct the model to draw inspiration from them to create a new Verilog module with similar code size and structural complexity but different functionality. We then discard newly generated modules that contain syntax errors.

\cu{(3) Spec generation.} Since GPT-4\cite{gpt4} has stronger capability to generate natural language than Claude-3.7, we use it to generate design specifications aligned with the newly produced Verilog modules. Same as common design practice in the hardware industry, we require each design specification to include: the module name; the functionality implemented by the module; a description of the input and output ports (the source text repeats “output”); a brief summary of the module’s high-level operation, including its basic operating principle, output behavior, critical combinational and timing behavior, and main control flow. To make the specification easier for the model to understand, we require the generated specification to be organized in a clear, hierarchical format using Markdown headers and bullet points.

\cu{(4) Filtering misaligned data.} Due to the hallucination of LLMs, the generated design specification may fail to accurately reflect the functionality and implementation logic of the code. To address this, we input both the generated design description and the code into Claude-3.7 and instruct the model to assume that the code is correct. Under this assumption, we let the model evaluate the quality of the design specification along four dimensions—accuracy, clarity, completeness, and conciseness, then output a final conclusion on whether the design specification is aligned with the code.

\cu{(5) Bug injection.} Based on our analysis of the benchmarks in representative researches on Verilog debugging including Strider\cite{strider}, MEIC\cite{meic}, Cirfix\cite{cirfix}, RTLRepair\cite{rtlrepair} and a proprietary industrial dataset from our industry collaborator, we inject bugs into correct code to obtain the buggy code according to five common Verilog functional bug patterns:

\begin{enumerate}
    \item Operator errors. Misuse of operators. Example: using arithmetic addition (+) instead of bitwise OR (|) to combine two flag vectors.
    \item Numerical value errors. Incorrect numerical values or bit widths. Example: "data\_out = 8'h1F;" instead of the intended "data\_out = 8'h2F;".
    \item Keyword errors. Use of an inappropriate keyword changes behavior. Example: confusing "input" and "output" reverses the signal direction.
    \item Variable name errors. Incorrect variable references. Example: using "count" in an expression when the intended signal is "cnt".
    \item Edge errors. Use of the wrong clock or reset edge. Example: "always @(negedge clk) q <= d;" instead of the intended "always @(posedge clk) q <= d;".
\end{enumerate}

We provide the design description, the correct code, and the specified bug type to Claude-3.7, and instruct it to inject exactly one bug of that type into the code without introducing any syntax error. We discard all generated variants that contain syntax errors or fail logic synthesis.

As shown in Table \ref{tab:datasetcom}, the pipeline above substantially enlarges our training dataset. In particular, to expand training data with relatively large modules, we apply the pipeline iteratively, in which newly generated modules with more than 200 lines from one iteration are fed back as seeds for the next iteration. The data synthesis terminates when all newly generated code duplicates existing code. 

\begin{table}[htbp]
\centering
\setlength{\tabcolsep}{5pt}
\caption{Comparison of datasets before and after expansion on sample count of each code scale.}
\begin{tabular}{cccc}
\hline
\textbf{Line counts} & \textbf{100$\sim$199} & \textbf{200$\sim$299} & \textbf{over 300} \\
\hline
Seed & 8687 & 781 & 217 \\
Enlarged & 14395 & 8633 & 2909\\
\hline
\end{tabular}
\label{tab:datasetcom}
\end{table}

\section{Evaluation}
\subsection{Experimental Setups} 
\subsubsection{Training Setups} We train the model using eight A800-80G GPUs with a learning rate of 1e-4. We apply cosine annealing for learning rate adjustment, and the first 10\% of training steps are used for warm-up.

\subsubsection{The Test Dataset}
\begin{figure}[t]
  \centering
\includegraphics[width=\linewidth]{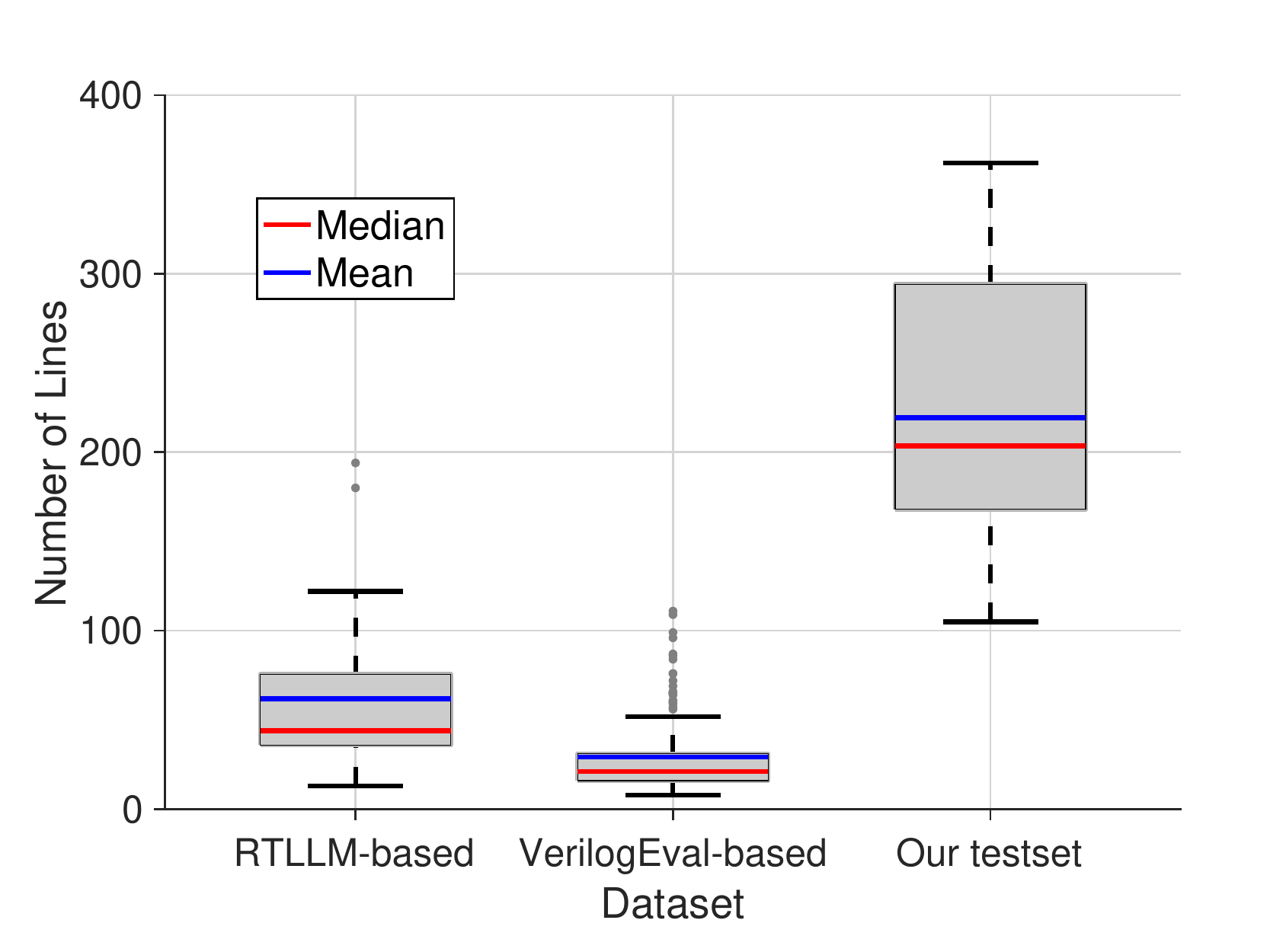}
  \caption{The code scale of test cases vs existing test sets based on RTLLM and VerilogEval.}\label{fig:scalecom}
  \vspace{-10pt}
\end{figure}

Existing open-source test datasets for LLM-based Verilog debugging are constructed from two representative benchmarks on the task of LLM-based Verilog generation, namely RTLLM\cite{rtllm} and VerilogEval \cite{verilogeval}. These code samples are small in scale (with RTLLM averaging 61 lines and VerilogEval averaging 29 lines), which can not reflect the complexity of real-world industrial design. To address this issue, we select two test sets for LLM Verilog coding-related tasks: the RTLLM and the recently proposed GenBen\cite{genben}. We choose them for their completeness and correctness: in both benchmarks, every sample contains a well-defined design specification, golden code well aligned with the design specification, and a testbench with complete functional coverage; all of these components have been rigorously reviewed by engineers and conform to industrial design standards. From these test sets, we extract 33 designs with more than 100 lines of code, which we use to construct our test dataset. Fig. \ref{fig:scalecom} presents a complexity comparison between our test set and existing RTLLM and VerilogEval-based test sets. Our test cases average 219 lines, significantly surpassing both RTLLM and VerilogEval-based benchmarks. 

To create buggy codes as test data, we systematically introduce bugs into the
golden code according to the common Verilog error types discussed
in Sec. \ref{sec:datasynthesis}. We ensure a balanced distribution across the bug categories. For each design, we create four buggy variants containing exactly 1, 2, 3, and 4 bugs respectively. Since we focus on functional bugs, all buggy variants are free of syntax errors, remain synthesizable, yet fail the corresponding testbench. The final test dataset contains 132 test cases. 

For each test case, we consider an output to be correct if the generated code passes the testbench.

\subsubsection{Evaluation Metrics} 
We use \xi{Pass@k} metric, which is widely We use the pass@k metric, which is widely employed in the evalution for coding-related tasks based on LLMs; it measures the probability of finding at least one correct answer within \xi{k} repeated generations, which can be estimated by: 
\begin{equation}
\xi{pass@k} := \mathbb{E}\left[ 1 - \frac{\binom{n-c}{k}}{\binom{n}{k}} \right]
\end{equation}
where \xi{n} denotes the number of generated outputs for each test case, and \xi{c} denotes the number of correct answers among \xi{n} generations. Consistent with prior studies on LLM-based coding and debugging\cite{uvllm,rtlcoder,rtlfixer,veriseek,verilogeval}, we set \xi{k} = 20 for reliable performance estimates. We report \xi{pass@1} and \xi{pass@5}, where \xi{pass@1} reflects the debugging accuracy under a single generated attempt, while \xi{pass@5} evaluates the performance when multiple repeated solutions are considered.

\subsubsection{Baseline Methods} 
We consider two categories of baselines: LLM-based and traditional non-LLM methods.

(1) LLM-based methods: Existing LLM-based Verilog debugging methods can be classified into two types: (a) Methods based on commercial LLMs, typically organized as a multi-agent Verilog debugging workflow; and (b) Methods based on open-source LLMs, where models are trained on public or internal datasets to enhance debugging performance. We take MEIC\cite{meic} and VeriDebug\cite{veridebug}, the SOTA methods of these two types as baselines. In addition, we compare our method against mainstream commercial LLMs with strong coding capabilities, including Claude-3.7, and DeepSeek-V3.

Restricted by their design, MEIC and VeriDebug must operate with a complete Verilog module. Commercial LLMs, however, are not subject to this constraint: they can always return an output for any input. Because our method follows a divide-and-conquer strategy, for fairness comparison, we evaluate two variants for the three commercial LLMs: (a) A default setting in which the complete code block is provided as a single input; and (b) A fragment-level setting same as ours, where we provide the semantically tight fragments derived from each test case, instruct the model to repair each fragment independently, and then reassemble the repaired fragments into the final fixed code.

(2) Traditional non-LLM methods. These approaches are based on fixed rules and templates. We employ the current SOTA method Strider\cite{strider} as a baseline. Strider first localizes the buggy line according to mismatched signals, and then generates patches with a signal value transition-guided approach. Since Strider relies on mismatched signals extracted from testbench outputs, we adjust the output format of the testbenches in the test set to match the input format required by Strider.

\subsection{Experimental Results}
\subsubsection{Repair Success Rate} 
Table \ref{tab:mainperformance} shows the overall repair success rate and the block-level success rate of our method, where Claude-3.7-r and Deepseek-3.7-r represent the method of letting the models repair each semantic fragment separately.

The experimental results demonstrate the efficiency of ARSP, achieving a pass@1 exceeding 75\% and a pass@5 exceeding 80\%. These results are attributable to two key strategies adopted by ARSP: semantic partitioning and training supported by high-quality synthetic data. Our test set exhibits substantially greater complexity than those in prior work on LLM-based Verilog debugging, whose test sets are based on RTLLM and VerilogEval with average line counts of 73 and 23 respectively. This indicates the potential of ARSP to assist the verification of industrial-scale modules.
\begin{table}[t]
	\centering
	\caption{Overall Repair Success Rate.}
	\begin{tabular}{ccccc}
		\toprule
		\textbf{Type} & \textbf{Method} & \textbf{pass@1(\%)} & \textbf{pass@5(\%)}\\
		\midrule
		\multirow{7}{*}{\makecell[c]{LLM}} &
            MEIC & 35.00 & 37.56 \\
            & Veridebug & 4.73 & 5.41 \\
            & Claude-3.7 & 40.15 & 52.42 \\
		& Claude-3.7-r  & 32.16 & 43.88  \\
            & Deepseek-V3 & 28.71 & 48.16  \\
		& Deepseek-V3-r & 22.95 & 35.00 \\
		\midrule
		\makecell[c]{non-LLM} & Strider & 28.03 & 28.03  \\
		\midrule
		Ours & ARSP & \cu{77.92} & \cu{83.88} \\
		\bottomrule
	\end{tabular}
	\vspace{0.5cm}
	\label{tab:mainperformance}
\end{table}

ARSP outperforms all baselines, exceeding the SOTA by over 35\% on pass@1 and 30\% on pass@5. Moreover, we observe that the performance of Claude and Deepseek decreases after applying semantic partition. This degradation arises because these models are general-purpose. During their training, the Verilog data they have seen predominantly reflects end-to-end fixes of entire code blocks rather than debugging at the level of individual semantic fragments. Since prior work did not adopt a fragmentation-and-local-repair paradigm, they have not been exposed to data following a divide-and-conquer debugging pipeline. Consequently, fragment-level prompts introduce a distribution shift that weakens their effectiveness.

Moreover, for Strider, since it is based on deterministic rules, it always produces the same answer for a given test case. Consequently, its \xi{pass@5} equals \xi{pass@1}. In our experiments, Strider attains a 56.06\% accuracy in bug localization, while its repair accuracy is only 28.03\%. This gap arises because bug localization relies largely on structural code analysis, which can be captured by rule-based heuristics. In contrast, producing correct repairs is a semantic task that requires understanding intended functionality and key signal relationships, which fixed rules cannot provide.
\begin{table*}[htbp]
\centering
\caption{Performance on different scales and bug numbers. The data in each cell is pass@1/pass@5.}
\begin{subtable}{\textwidth}
\centering
\setlength{\tabcolsep}{2.6pt}
\caption{Results on different code scales.}
\begin{tabular}{ccccccccc}
\hline
\textbf{Code scale} & \textbf{MEIC} & \textbf{Veridebug} & 
\textbf{Claude-3.7} &
\textbf{Claude-3.7-r} &
\textbf{Deepseek-V3} &
\textbf{Deepseek-V3-r} &
\textbf{Strider} &
\textbf{ARSP} \\
\hline
100$\sim$199 & 44.26/45.46 & 7.72/9.0 & 46.76/52.67 & 37.13/44.56 & 35.51/50.5 & 32.2/44.55 & 30.88/30.88 & \cu{82.64/86.45}\\
200$\sim$299 & 32.91/33.28 & 0/0 & 40.94/58.4 & 26.04/41.03 & 25.72/50.75 & 16.46/29.98 & 31.25/31.25 & \cu{75.52/83.37}\\
over 300 & 1.88/3.58 & 6.25/6.25 & 9.69/33.4 & 29.38/49.54 & 8.75/30.41 & 3.13/9.49 & 6.25/6.25 & \cu{65.00/74.47}\\
\hline
\end{tabular}
\label{tab:codescale}
\end{subtable}
\vspace{1em}
\begin{subtable}{\textwidth}
\centering
\setlength{\tabcolsep}{3pt}
\caption{Results on different bug numbers.}
\begin{tabular}{ccccccccc}
\hline
\textbf{\# of bugs} & \textbf{MEIC} & \textbf{Veridebug} & 
\textbf{Claude-3.7} &
\textbf{Claude-3.7-r} &
\textbf{Deepseek-V3} &
\textbf{Deepseek-V3-r} &
\textbf{Strider} &
\textbf{ARSP} \\
\hline
1 & 39.39/43.33 & 9.54/11.2 & 39.55/53.3 & 35.9/49.3 & 31.51/53.8 & 22.12/39.89 & 45.45/45.45 & \cu{85.75/90.69}\\
2 & 35.58/36.29 & 3.23/4.25 & 41.03/52.11 & 33.23/45.91 & 26.17/44.74 & 30.88/39.93 & 26.47/26.47 & \cu{80.73/86.21}\\
3 & 33.79/35.42 & 3.03/3.03 & 38.64/51.86 & 31.51/41.77 & 29.39/49.93 & 18.79/31.09 & 21.21/21.21 & \cu{76.06/81.51}\\
4 & 31.09/35.16 & 3.13/3.13 & 41.4/52.39 & 27.81/38.32 & 27.92/44.15 & 19.69/28.73 & 18.75/18.75 & \cu{68.75/76.81}\\
\hline
\end{tabular}
\label{tab:bugnum}
\end{subtable}
\label{tab:scaleperformance}
\end{table*}
We observe that Veridebug exhibits low performance. This is because its training set is simple, in which each sample contains only one bug and has an average length of 45 lines. This is substantially below the complexity of our test dataset.

\subsubsection{Results on Different Code Scales and Number of bugs}

Table \ref{tab:scaleperformance} reports the performance across varying code scales and numbers of Bugs in the test case. As code length and the number of bugs increase, the difficulty of the cases rises and overall performance degrades. ARSP outperforms all baselines across all code sizes and bug counts, with a particularly substantial advantage when the code scale exceeds 300 lines. This is because our semantic partition strategy explicitly targets the problem of bug signal dilution within the broader context of longer code, and our data generation pipeline mainly yields longer code samples. These factors together give ARSP a comparative advantage in processing larger Verilog modules.

\subsubsection{Performance of the Partition LLM} On our test dataset, Partition LLM splits each module into 5.73 fragments on average, with an average fragment length of 41.3 lines. This substantially reduces the debugging context and mitigates the dilution of bug signals in long-code contexts. 

The pass@1 success rate of Partition LLM’s semantic partitioning is 96.51\%; after applying the rule-based correction method described in Sec. \ref{sec:rulebasedcorrection}, the success rate further increases to 99.24\%. The occasional partition failures are discussed in Sec. \ref{sec:partitionfail}.

\subsubsection{Ablation Studies}
To assess the impact of the semantic partition strategy and the newly synthesized data on ARSP’s performance, we evaluated the following three models' debugging success rate on our test dataset:

(1) Base model.

(2) The Non-partition Repair LLM trained only on the seed dataset, which is denoted by \xi{NP$_s$}.

(3) The Non-partition Repair LLM trained on both the seed and newly generated dataset, representied by \xi{NP$_{s+g}$}. This is the Non-partition LLM used in the ARSP workflow.

The results are shown in Table. \ref{tab:ablation}. Firstly, since the base model itself has not received task-specific supervision for Verilog debugging, its performance is pretty low. Relative to the base model, the supervised Non-partition LLM \xi{NP$_{s}$} exhibits a certain level of debugging capability. In comparison, after supervised training on aligned buggy–correct code pairs, \xi{NP$_s$} exhibits some debugging capability. Then, after training with newly generated data, \xi{NP$_{s+g}$} shows improved performance over \xi{NP$_{s}$}. 

A substantial improvement is observed when applying the semantic partition: instead of the non-partition LLM \xi{NP$_{s+g}$} processing an entire Verilog module in a single pass, ARSP splits the module into multiple semantic tight fragments and debugs them individually. This increases pass@1 by 11.6\% and pass@5 by 10.2\%. This comparison is fair for the following reasons. Our semantic partition strategy depends on two components: the Partition LLM and the Repair LLM. The data required to train them is obtained solely by splitting the buggy and correct code in each training example of the Non-partition Repair LLM; no additional module code is generated and no new bugs are inserted. 

\begin{table}[t]
\centering
\caption{Ablation Studies.}
\begin{subtable}{\columnwidth}
\centering
\setlength{\tabcolsep}{5pt}
\caption{Overall repair success rate.}
\begin{tabular}{ccc}
\hline
\textbf{Method} & \textbf{pass@1} & \textbf{pass@5} \\
\hline
Base model & 4.09 & 9.18\\
\xi{NP$_s$} & 60.6  &  63.94\\
\xi{NP$_{s+g}$} & 66.32 & 73.63\\
ARSP & 77.92 & 83.88\\
\hline
\end{tabular}
\label{tab:ablationmain}
\end{subtable}
\vspace{-2pt}

\begin{subtable}{\columnwidth}
\centering
\setlength{\tabcolsep}{2.3pt}
\caption{Results on different code scales.}
\small
\begin{tabular}{ccccc}
\hline
\textbf{Code scale} & \textbf{Base} & \textbf{\xi{NP$_s$}} & 
\textbf{\xi{NP$_{s+g}$}} &
\textbf{ARSP} \\
\hline
100$\sim$199 & 5.22/12.26 & 70.22/73.86 & 76.1/84.79 & \cu{82.64/86.45} \\
200$\sim$299 & 3.85/7.88 & 60.42/63.69 & 66.88/71.99 & \cu{75.52/83.37} \\
over 300 & 0/0 & 20.31/22.5 & 23.13/31.15 & \cu{65.0/74.47} \\
\hline
\end{tabular}
\label{tab:ablationscale}
\end{subtable}
\label{tab:ablation}
\end{table}
More importantly, as shown in Table \ref{tab:ablationscale}, for test cases exceeding 300 lines, semantic partition yields a 41.9\% improvement in pass@1 and a 43.32\% improvement in pass@5 without introducing new module code or inserting new bugs. This far exceeds the gains from data synthesis, in which pass@1 and pass@5 increase by only 2.82\% and 8.65\% respectively. This demonstrates that addressing the dilution of bug signals in long-code contexts is crucial for LLM-based Verilog debugging, which is no less important than obtaining sufficient training data.


\section{Discussion}
\subsection{About New Bug Types}
The bug categories considered in this paper are derived by two engineers with five years of industrial experience through analysis of the benchmarks used in existing researches on Verilog debugging and 124 historical Verilog patches obtained from our industry partner. These categories are intended to capture the most representative and frequently occurring functional error patterns in practical Verilog designs, rather than serving as an exhaustive list of all possible bug types. We acknowledge that bugs introduced by human engineers are virtually endless, and new bug types may also emerge in the future. Nevertheless, this does not affect the practical utility of our method. ARSP is inherently bug-agnostic. In practice, by summarizing new bug types and using the data synthesis pipeline in Sec. \ref{sec:datasynthesis}, we can create more training data for these new bug types, enabling the model to address them. This is similar to the industrial practice of maintaining historical bug repositories: companies such as ARM and Intel maintain repositories documenting bugs encountered in prior projects to guide current designs. When new bug categories are discovered, these repositories are updated accordingly. Likewise, ARSP can be continually improved by updating its training datasets, thereby maintaining adaptability to future design challenges.

\subsection{Occasional Partition Failures}\label{sec:partitionfail}
Due to the inherent randomness of LLM outputs, the fragments produced by the Partition LLM may not always recombine into the original buggy code perfectly. In most such cases, the total number of lines after concatenation remains identical to that of the original buggy code, allowing us to apply the rule-based method to correct it as described in Sec. \ref{sec:rulebasedcorrection}. In other cases, however, as we have observed, the errors in the fragments exhibit greater uncertainty. For example, some lines are missing at arbitrary positions, portions of code are collapsed into "...", or the order of certain lines is altered. This uncertainty makes it difficult to correct them with a determined method. Although the success rate of partitioning has already reached 99\%, for completeness, we will consider improved solutions to this problem in future work.

\section{Related Works}
Given that Verilog debugging is a time-consuming task, automated debugging techniques have received much attention in recent years. Existing research can be mainly divided into two categories: rule and template-driven methods \cite{cirfix,asplos22,rtlrepair,aspdac24,strider}, and LLM-based methods \cite{TCS,TIFS24,uvllm,veriassist,sp23,todaes25,rtlfixer,aivril,veridebug,meic,hdldebugger}. Regarding the rule and template-driven approach, \cite{cirfix} proposes the first automated end-to-end debugging framework Cirfix for hardware description languages. \cite{asplos22} combines static analysis and dynamic verification to construct a workflow for debugging FPGA designs. \cite{aspdac24} reduces redundant simulation by injecting new logic into existing waveforms. \cite{rtlrepair} proposes a symbolic, template-based method that uses SMT solvers for bug localization and repair. \cite{strider} detects bugs through analysis of mismatched signals and generates patches based on expected signal transitions. These methods rely on predefined rules and templates and lack understanding of high-level functional semantics and design specifications, which limits their effectiveness in scenarios with diverse functional bugs.

Inspired by strong code understanding capabilities of LLMs, recent research has explored LLM-based automated debugging methods for Verilog. \cite{sp23,TIFS24} investigated prompt engineering techniques for repairing security vulnerabilities in Verilog. HDLdebugger \cite{hdldebugger,TCS} proposed training datasets on the verification for RTL code verification tasks. UVLLM\cite{uvllm} introduced a Universal Verification Methodology (UVM) framework based on LLMs. VeriAssist \cite{veriassist} and AIvril \cite{aivril} proposed multi-agent systems that integrate code generation, verification scripts generation, and bug repair. \cite{todaes25} presented a etrieval-Augmented-Generation (RAG) based framework to improve repair accuracy. RTL-Fixer \cite{rtlfixer} developed a workflow based on the prompting strategy of ReAct\cite{ReAct} to fix syntax errors. \cite{veridebug} proposed an open-source unified LLM for bug localization, classification and repair. MEIC \cite{meic} proposed an iterative framework that advances the repair process via a code scoring mechanism. However, none of these works address the bug signal dilution caused by long code contexts.

\section{Conclusion}
In this paper, we propose ARSP, an automated Verilog debugging system based on semantic partitioning. To address the problem of bug signal dilution in long contexts of industrial-scale Verilog designs, ARSP employs a divide-and-conquer strategy that first partitions buggy code into semantically tight fragments, then debugs each fragment separately, and finally merges them into the repaired code. To overcome the scarcity of training data for Verilog debugging, we develope a data synthesis framework to create high-quality data samples with aligned design specification, correct and buggy code pairs. Experimental results demonstrate the superior performance of ARSP, significantly outperforming existing commercial LLMs and SOTA automated Verilog debugging tools. Notably, semantic partitioning alone contributes significant improvements of 11.6\% in pass@1 and 10.2\% in pass@5 over whole-module debugging without introducing new training data. This offers a new perspective for improving LLM's performance on Verilog debugging beyond collecting more training data.

\bibliographystyle{ACM-Reference-Format}
\bibliography{sample-base}


\begin{thebibliography}{42}


\ifx \showCODEN    \undefined \def \showCODEN     #1{\unskip}     \fi
\ifx \showDOI      \undefined \def \showDOI       #1{#1}\fi
\ifx \showISBNx    \undefined \def \showISBNx     #1{\unskip}     \fi
\ifx \showISBNxiii \undefined \def \showISBNxiii  #1{\unskip}     \fi
\ifx \showISSN     \undefined \def \showISSN      #1{\unskip}     \fi
\ifx \showLCCN     \undefined \def \showLCCN      #1{\unskip}     \fi
\ifx \shownote     \undefined \def \shownote      #1{#1}          \fi
\ifx \showarticletitle \undefined \def \showarticletitle #1{#1}   \fi
\ifx \showURL      \undefined \def \showURL       {\relax}        \fi
\providecommand\bibfield[2]{#2}
\providecommand\bibinfo[2]{#2}
\providecommand\natexlab[1]{#1}
\providecommand\showeprint[2][]{arXiv:#2}

\bibitem[dee(2023)]%
        {deepseekcoder}
 \bibinfo{year}{2023}\natexlab{}.
\newblock \bibinfo{title}{DeepSeek-Coder-6.7B-Base}.
\newblock \bibinfo{howpublished}{\url{https://huggingface.co/deepseek-ai/deepseek-coder-6.7b-base}}.
\newblock


\bibitem[cla(2024)]%
        {claudeapi}
 \bibinfo{year}{2024}\natexlab{}.
\newblock \bibinfo{title}{Claude API overview}.
\newblock \bibinfo{howpublished}{\url{https://docs.anthropic.com/en/docs/about-claude/models/overview}}.
\newblock


\bibitem[gpt(2024)]%
        {gpt4}
 \bibinfo{year}{2024}\natexlab{}.
\newblock \bibinfo{title}{GPT-4}.
\newblock \bibinfo{howpublished}{\url{https://platform.openai.com/docs/api-reference/introduction}}.
\newblock


\bibitem[Ahmad et~al\mbox{.}(2024)]%
        {TIFS24}
\bibfield{author}{\bibinfo{person}{Baleegh Ahmad}, \bibinfo{person}{Shailja Thakur}, \bibinfo{person}{Benjamin Tan}, \bibinfo{person}{Ramesh Karri}, {and} \bibinfo{person}{Hammond Pearce}.} \bibinfo{year}{2024}\natexlab{}.
\newblock \showarticletitle{On Hardware Security Bug Code Fixes by Prompting Large Language Models}.
\newblock \bibinfo{journal}{\emph{IEEE Transactions on Information Forensics and Security}}  \bibinfo{volume}{19} (\bibinfo{year}{2024}), \bibinfo{pages}{4043--4057}.
\newblock
\urldef\tempurl%
\url{https://doi.org/10.1109/TIFS.2024.3374558}
\showDOI{\tempurl}


\bibitem[Fu et~al\mbox{.}(2025)]%
        {TCS}
\bibfield{author}{\bibinfo{person}{Weimin Fu}, \bibinfo{person}{Shijie Li}, \bibinfo{person}{Yifang Zhao}, \bibinfo{person}{Kaichen Yang}, \bibinfo{person}{Xuan Zhang}, \bibinfo{person}{Yier Jin}, {and} \bibinfo{person}{Xiaolong Guo}.} \bibinfo{year}{2025}\natexlab{}.
\newblock \showarticletitle{A Generalize Hardware Debugging Approach for Large Language Models Semi-Synthetic, Datasets}.
\newblock \bibinfo{journal}{\emph{IEEE Transactions on Circuits and Systems I: Regular Papers}} \bibinfo{volume}{72}, \bibinfo{number}{2} (\bibinfo{year}{2025}), \bibinfo{pages}{623--636}.
\newblock
\urldef\tempurl%
\url{https://doi.org/10.1109/TCSI.2024.3487486}
\showDOI{\tempurl}


\bibitem[Gu et~al\mbox{.}(2025)]%
        {llmasjudge1}
\bibfield{author}{\bibinfo{person}{Jiawei Gu}, \bibinfo{person}{Xuhui Jiang}, \bibinfo{person}{Zhichao Shi}, \bibinfo{person}{Hexiang Tan}, \bibinfo{person}{Xuehao Zhai}, \bibinfo{person}{Chengjin Xu}, \bibinfo{person}{Wei Li}, \bibinfo{person}{Yinghan Shen}, \bibinfo{person}{Shengjie Ma}, \bibinfo{person}{Honghao Liu}, \bibinfo{person}{Saizhuo Wang}, \bibinfo{person}{Kun Zhang}, \bibinfo{person}{Yuanzhuo Wang}, \bibinfo{person}{Wen Gao}, \bibinfo{person}{Lionel Ni}, {and} \bibinfo{person}{Jian Guo}.} \bibinfo{year}{2025}\natexlab{}.
\newblock \bibinfo{title}{A Survey on LLM-as-a-Judge}.
\newblock
\newblock
\showeprint[arxiv]{2411.15594}~[cs.CL]
\urldef\tempurl%
\url{https://arxiv.org/abs/2411.15594}
\showURL{%
\tempurl}


\bibitem[Hu et~al\mbox{.}(2022)]%
        {lora}
\bibfield{author}{\bibinfo{person}{Edward~J Hu}, \bibinfo{person}{yelong shen}, \bibinfo{person}{Phillip Wallis}, \bibinfo{person}{Zeyuan Allen-Zhu}, \bibinfo{person}{Yuanzhi Li}, \bibinfo{person}{Shean Wang}, \bibinfo{person}{Lu Wang}, {and} \bibinfo{person}{Weizhu Chen}.} \bibinfo{year}{2022}\natexlab{}.
\newblock \showarticletitle{Lo{RA}: Low-Rank Adaptation of Large Language Models}. In \bibinfo{booktitle}{\emph{International Conference on Learning Representations}}.
\newblock
\urldef\tempurl%
\url{https://openreview.net/forum?id=nZeVKeeFYf9}
\showURL{%
\tempurl}


\bibitem[Hu et~al\mbox{.}(2024)]%
        {uvllm}
\bibfield{author}{\bibinfo{person}{Yuchen Hu}, \bibinfo{person}{Junhao Ye}, \bibinfo{person}{Ke Xu}, \bibinfo{person}{Jialin Sun}, \bibinfo{person}{Shiyue Zhang}, \bibinfo{person}{Xinyao Jiao}, \bibinfo{person}{Dingrong Pan}, \bibinfo{person}{Jie Zhou}, \bibinfo{person}{Ning Wang}, \bibinfo{person}{Weiwei Shan}, \bibinfo{person}{Xinwei Fang}, \bibinfo{person}{Xi Wang}, \bibinfo{person}{Nan Guan}, {and} \bibinfo{person}{Zhe Jiang}.} \bibinfo{year}{2024}\natexlab{}.
\newblock \bibinfo{title}{UVLLM: An Automated Universal RTL Verification Framework using LLMs}.
\newblock
\newblock
\showeprint[arxiv]{2411.16238}~[cs.AR]
\urldef\tempurl%
\url{https://arxiv.org/abs/2411.16238}
\showURL{%
\tempurl}


\bibitem[Huang et~al\mbox{.}(2024)]%
        {veriassist}
\bibfield{author}{\bibinfo{person}{Hanxian Huang}, \bibinfo{person}{Zhenghan Lin}, \bibinfo{person}{Zixuan Wang}, \bibinfo{person}{Xin Chen}, \bibinfo{person}{Ke Ding}, {and} \bibinfo{person}{Jishen Zhao}.} \bibinfo{year}{2024}\natexlab{}.
\newblock \bibinfo{title}{Towards LLM-Powered Verilog RTL Assistant: Self-Verification and Self-Correction}.
\newblock
\newblock
\showeprint[arxiv]{2406.00115}~[cs.PL]
\urldef\tempurl%
\url{https://arxiv.org/abs/2406.00115}
\showURL{%
\tempurl}


\bibitem[Husain et~al\mbox{.}(2020)]%
        {minhash}
\bibfield{author}{\bibinfo{person}{Hamel Husain}, \bibinfo{person}{Ho-Hsiang Wu}, \bibinfo{person}{Tiferet Gazit}, \bibinfo{person}{Miltiadis Allamanis}, {and} \bibinfo{person}{Marc Brockschmidt}.} \bibinfo{year}{2020}\natexlab{}.
\newblock \bibinfo{title}{CodeSearchNet Challenge: Evaluating the State of Semantic Code Search}.
\newblock
\newblock
\showeprint[arxiv]{1909.09436}~[cs.LG]
\urldef\tempurl%
\url{https://arxiv.org/abs/1909.09436}
\showURL{%
\tempurl}


\bibitem[Jafari et~al\mbox{.}(2021)]%
        {lsh}
\bibfield{author}{\bibinfo{person}{Omid Jafari}, \bibinfo{person}{Preeti Maurya}, \bibinfo{person}{Parth Nagarkar}, \bibinfo{person}{Khandker~Mushfiqul Islam}, {and} \bibinfo{person}{Chidambaram Crushev}.} \bibinfo{year}{2021}\natexlab{}.
\newblock \bibinfo{title}{A Survey on Locality Sensitive Hashing Algorithms and their Applications}.
\newblock
\newblock
\showeprint[arxiv]{2102.08942}~[cs.DB]
\urldef\tempurl%
\url{https://arxiv.org/abs/2102.08942}
\showURL{%
\tempurl}


\bibitem[Jiang et~al\mbox{.}(2024)]%
        {llmcoding2}
\bibfield{author}{\bibinfo{person}{Juyong Jiang}, \bibinfo{person}{Fan Wang}, \bibinfo{person}{Jiasi Shen}, \bibinfo{person}{Sungju Kim}, {and} \bibinfo{person}{Sunghun Kim}.} \bibinfo{year}{2024}\natexlab{}.
\newblock \bibinfo{title}{A Survey on Large Language Models for Code Generation}.
\newblock
\newblock
\showeprint[arxiv]{2406.00515}~[cs.CL]
\urldef\tempurl%
\url{https://arxiv.org/abs/2406.00515}
\showURL{%
\tempurl}


\bibitem[Klemmer and Große(2024)]%
        {aspdac24}
\bibfield{author}{\bibinfo{person}{Lucas Klemmer} {and} \bibinfo{person}{Daniel Große}.} \bibinfo{year}{2024}\natexlab{}.
\newblock \showarticletitle{Towards a Highly Interactive Design-Debug-Verification Cycle}. In \bibinfo{booktitle}{\emph{2024 29th Asia and South Pacific Design Automation Conference (ASP-DAC)}}. \bibinfo{pages}{692--697}.
\newblock
\urldef\tempurl%
\url{https://doi.org/10.1109/ASP-DAC58780.2024.10473953}
\showDOI{\tempurl}


\bibitem[Laeufer et~al\mbox{.}(2024)]%
        {rtlrepair}
\bibfield{author}{\bibinfo{person}{Kevin Laeufer}, \bibinfo{person}{Brandon Fajardo}, \bibinfo{person}{Abhik Ahuja}, \bibinfo{person}{Vighnesh Iyer}, \bibinfo{person}{Borivoje Nikoli\'{c}}, {and} \bibinfo{person}{Koushik Sen}.} \bibinfo{year}{2024}\natexlab{}.
\newblock \showarticletitle{RTL-Repair: Fast Symbolic Repair of Hardware Design Code}. In \bibinfo{booktitle}{\emph{Proceedings of the 29th ACM International Conference on Architectural Support for Programming Languages and Operating Systems, Volume 3}} (La Jolla, CA, USA) \emph{(\bibinfo{series}{ASPLOS '24})}. \bibinfo{publisher}{Association for Computing Machinery}, \bibinfo{address}{New York, NY, USA}, \bibinfo{pages}{867–881}.
\newblock
\showISBNx{9798400703867}
\urldef\tempurl%
\url{https://doi.org/10.1145/3620666.3651346}
\showDOI{\tempurl}


\bibitem[Lahti et~al\mbox{.}(2019)]%
        {debuggingtime}
\bibfield{author}{\bibinfo{person}{Sakari Lahti}, \bibinfo{person}{Panu Sjövall}, \bibinfo{person}{Jarno Vanne}, {and} \bibinfo{person}{Timo~D. Hämäläinen}.} \bibinfo{year}{2019}\natexlab{}.
\newblock \showarticletitle{Are We There Yet? A Study on the State of High-Level Synthesis}.
\newblock \bibinfo{journal}{\emph{IEEE Transactions on Computer-Aided Design of Integrated Circuits and Systems}} \bibinfo{volume}{38}, \bibinfo{number}{5} (\bibinfo{year}{2019}), \bibinfo{pages}{898--911}.
\newblock
\urldef\tempurl%
\url{https://doi.org/10.1109/TCAD.2018.2834439}
\showDOI{\tempurl}


\bibitem[Liu et~al\mbox{.}(2023)]%
        {verilogeval}
\bibfield{author}{\bibinfo{person}{Mingjie Liu}, \bibinfo{person}{Nathaniel Pinckney}, \bibinfo{person}{Brucek Khailany}, {and} \bibinfo{person}{Haoxing Ren}.} \bibinfo{year}{2023}\natexlab{}.
\newblock \showarticletitle{VerilogEval: Evaluating Large Language Models for Verilog Code Generation}. In \bibinfo{booktitle}{\emph{2023 IEEE/ACM International Conference on Computer Aided Design (ICCAD)}}. \bibinfo{pages}{1--8}.
\newblock
\urldef\tempurl%
\url{https://doi.org/10.1109/ICCAD57390.2023.10323812}
\showDOI{\tempurl}


\bibitem[Liu et~al\mbox{.}(2025)]%
        {rtlcoder}
\bibfield{author}{\bibinfo{person}{Shang Liu}, \bibinfo{person}{Wenji Fang}, \bibinfo{person}{Yao Lu}, \bibinfo{person}{Jing Wang}, \bibinfo{person}{Qijun Zhang}, \bibinfo{person}{Hongce Zhang}, {and} \bibinfo{person}{Zhiyao Xie}.} \bibinfo{year}{2025}\natexlab{}.
\newblock \showarticletitle{RTLCoder: Fully Open-Source and Efficient LLM-Assisted RTL Code Generation Technique}.
\newblock \bibinfo{journal}{\emph{IEEE Transactions on Computer-Aided Design of Integrated Circuits and Systems}} \bibinfo{volume}{44}, \bibinfo{number}{4} (\bibinfo{year}{2025}), \bibinfo{pages}{1448--1461}.
\newblock
\urldef\tempurl%
\url{https://doi.org/10.1109/TCAD.2024.3483089}
\showDOI{\tempurl}


\bibitem[Lu et~al\mbox{.}(2024)]%
        {rtllm}
\bibfield{author}{\bibinfo{person}{Yao Lu}, \bibinfo{person}{Shang Liu}, \bibinfo{person}{Qijun Zhang}, {and} \bibinfo{person}{Zhiyao Xie}.} \bibinfo{year}{2024}\natexlab{}.
\newblock \showarticletitle{RTLLM: An Open-Source Benchmark for Design RTL Generation with Large Language Model}. In \bibinfo{booktitle}{\emph{2024 29th Asia and South Pacific Design Automation Conference (ASP-DAC)}}. \bibinfo{pages}{722--727}.
\newblock
\urldef\tempurl%
\url{https://doi.org/10.1109/ASP-DAC58780.2024.10473904}
\showDOI{\tempurl}


\bibitem[Ma et~al\mbox{.}(2022)]%
        {asplos22}
\bibfield{author}{\bibinfo{person}{Jiacheng Ma}, \bibinfo{person}{Gefei Zuo}, \bibinfo{person}{Kevin Loughlin}, \bibinfo{person}{Haoyang Zhang}, \bibinfo{person}{Andrew Quinn}, {and} \bibinfo{person}{Baris Kasikci}.} \bibinfo{year}{2022}\natexlab{}.
\newblock \showarticletitle{Debugging in the brave new world of reconfigurable hardware}. In \bibinfo{booktitle}{\emph{Proceedings of the 27th ACM International Conference on Architectural Support for Programming Languages and Operating Systems}} (Lausanne, Switzerland) \emph{(\bibinfo{series}{ASPLOS '22})}. \bibinfo{publisher}{Association for Computing Machinery}, \bibinfo{address}{New York, NY, USA}, \bibinfo{pages}{946–962}.
\newblock
\showISBNx{9781450392051}
\urldef\tempurl%
\url{https://doi.org/10.1145/3503222.3507701}
\showDOI{\tempurl}


\bibitem[McHugh(2012)]%
        {cohenkappa}
\bibfield{author}{\bibinfo{person}{Mary~L McHugh}.} \bibinfo{year}{2012}\natexlab{}.
\newblock \showarticletitle{Interrater reliability: the kappa statistic}.
\newblock \bibinfo{journal}{\emph{Biochemia medica}} \bibinfo{volume}{22}, \bibinfo{number}{3} (\bibinfo{year}{2012}), \bibinfo{pages}{276--282}.
\newblock


\bibitem[Minaee et~al\mbox{.}(2025)]%
        {llmsurvey}
\bibfield{author}{\bibinfo{person}{Shervin Minaee}, \bibinfo{person}{Tomas Mikolov}, \bibinfo{person}{Narjes Nikzad}, \bibinfo{person}{Meysam Chenaghlu}, \bibinfo{person}{Richard Socher}, \bibinfo{person}{Xavier Amatriain}, {and} \bibinfo{person}{Jianfeng Gao}.} \bibinfo{year}{2025}\natexlab{}.
\newblock \bibinfo{title}{Large Language Models: A Survey}.
\newblock
\newblock
\showeprint[arxiv]{2402.06196}~[cs.CL]
\urldef\tempurl%
\url{https://arxiv.org/abs/2402.06196}
\showURL{%
\tempurl}


\bibitem[Nadimi et~al\mbox{.}(2025)]%
        {pyranet}
\bibfield{author}{\bibinfo{person}{Bardia Nadimi}, \bibinfo{person}{Ghali~Omar Boutaib}, {and} \bibinfo{person}{Hao Zheng}.} \bibinfo{year}{2025}\natexlab{}.
\newblock \bibinfo{title}{PyraNet: A Multi-Layered Hierarchical Dataset for Verilog}.
\newblock
\newblock
\showeprint[arxiv]{2412.06947}~[cs.AR]
\urldef\tempurl%
\url{https://arxiv.org/abs/2412.06947}
\showURL{%
\tempurl}


\bibitem[Nam et~al\mbox{.}(2024)]%
        {llmcoding}
\bibfield{author}{\bibinfo{person}{Daye Nam}, \bibinfo{person}{Andrew Macvean}, \bibinfo{person}{Vincent Hellendoorn}, \bibinfo{person}{Bogdan Vasilescu}, {and} \bibinfo{person}{Brad Myers}.} \bibinfo{year}{2024}\natexlab{}.
\newblock \showarticletitle{Using an LLM to Help With Code Understanding}. In \bibinfo{booktitle}{\emph{Proceedings of the IEEE/ACM 46th International Conference on Software Engineering}} (Lisbon, Portugal) \emph{(\bibinfo{series}{ICSE '24})}. \bibinfo{publisher}{Association for Computing Machinery}, \bibinfo{address}{New York, NY, USA}, Article \bibinfo{articleno}{97}, \bibinfo{numpages}{13}~pages.
\newblock
\showISBNx{9798400702174}
\urldef\tempurl%
\url{https://doi.org/10.1145/3597503.3639187}
\showDOI{\tempurl}


\bibitem[Pearce et~al\mbox{.}(2023)]%
        {sp23}
\bibfield{author}{\bibinfo{person}{Hammond Pearce}, \bibinfo{person}{Benjamin Tan}, \bibinfo{person}{Baleegh Ahmad}, \bibinfo{person}{Ramesh Karri}, {and} \bibinfo{person}{Brendan Dolan-Gavitt}.} \bibinfo{year}{2023}\natexlab{}.
\newblock \showarticletitle{{ Examining Zero-Shot Vulnerability Repair with Large Language Models }}. In \bibinfo{booktitle}{\emph{2023 IEEE Symposium on Security and Privacy (SP)}}. \bibinfo{publisher}{IEEE Computer Society}, \bibinfo{address}{Los Alamitos, CA, USA}, \bibinfo{pages}{2339--2356}.
\newblock
\urldef\tempurl%
\url{https://doi.org/10.1109/SP46215.2023.10179420}
\showDOI{\tempurl}


\bibitem[Qayyum et~al\mbox{.}(2025)]%
        {todaes25}
\bibfield{author}{\bibinfo{person}{Khushboo Qayyum}, \bibinfo{person}{Chandan~Kumar Jha}, \bibinfo{person}{Sallar Ahmadi-Pour}, \bibinfo{person}{Muhammad Hassan}, {and} \bibinfo{person}{Rolf Drechsler}.} \bibinfo{year}{2025}\natexlab{}.
\newblock \showarticletitle{LLM-assisted Bug Identification and Correction for Verilog HDL}.
\newblock \bibinfo{journal}{\emph{ACM Trans. Des. Autom. Electron. Syst.}} (\bibinfo{date}{May} \bibinfo{year}{2025}).
\newblock
\showISSN{1084-4309}
\urldef\tempurl%
\url{https://doi.org/10.1145/3733237}
\showDOI{\tempurl}
\newblock
\shownote{Just Accepted}.


\bibitem[Santiesteban et~al\mbox{.}(2023)]%
        {cirfix}
\bibfield{author}{\bibinfo{person}{Priscila Santiesteban}, \bibinfo{person}{Yu Huang}, \bibinfo{person}{Westley Weimer}, {and} \bibinfo{person}{Hammad Ahmad}.} \bibinfo{year}{2023}\natexlab{}.
\newblock \showarticletitle{CirFix: Automated Hardware Repair and its Real-World Applications}.
\newblock \bibinfo{journal}{\emph{IEEE Transactions on Software Engineering}} \bibinfo{volume}{49}, \bibinfo{number}{7} (\bibinfo{year}{2023}), \bibinfo{pages}{3736--3752}.
\newblock
\urldef\tempurl%
\url{https://doi.org/10.1109/TSE.2023.3269899}
\showDOI{\tempurl}


\bibitem[Thakur et~al\mbox{.}(2024)]%
        {verigen}
\bibfield{author}{\bibinfo{person}{Shailja Thakur}, \bibinfo{person}{Baleegh Ahmad}, \bibinfo{person}{Hammond Pearce}, \bibinfo{person}{Benjamin Tan}, \bibinfo{person}{Brendan Dolan-Gavitt}, \bibinfo{person}{Ramesh Karri}, {and} \bibinfo{person}{Siddharth Garg}.} \bibinfo{year}{2024}\natexlab{}.
\newblock \showarticletitle{VeriGen: A Large Language Model for Verilog Code Generation}.
\newblock \bibinfo{journal}{\emph{ACM Trans. Des. Autom. Electron. Syst.}} \bibinfo{volume}{29}, \bibinfo{number}{3}, Article \bibinfo{articleno}{46} (\bibinfo{date}{April} \bibinfo{year}{2024}), \bibinfo{numpages}{31}~pages.
\newblock
\showISSN{1084-4309}
\urldef\tempurl%
\url{https://doi.org/10.1145/3643681}
\showDOI{\tempurl}


\bibitem[Tsai et~al\mbox{.}(2024)]%
        {rtlfixer}
\bibfield{author}{\bibinfo{person}{Yunda Tsai}, \bibinfo{person}{Mingjie Liu}, {and} \bibinfo{person}{Haoxing Ren}.} \bibinfo{year}{2024}\natexlab{}.
\newblock \showarticletitle{RTLFixer: Automatically Fixing RTL Syntax Errors with Large Language Model}. In \bibinfo{booktitle}{\emph{Proceedings of the 61st ACM/IEEE Design Automation Conference}} (San Francisco, CA, USA) \emph{(\bibinfo{series}{DAC '24})}. \bibinfo{publisher}{Association for Computing Machinery}, \bibinfo{address}{New York, NY, USA}, Article \bibinfo{articleno}{53}, \bibinfo{numpages}{6}~pages.
\newblock
\showISBNx{9798400706011}
\urldef\tempurl%
\url{https://doi.org/10.1145/3649329.3657353}
\showDOI{\tempurl}


\bibitem[ul~Islam et~al\mbox{.}(2024)]%
        {aivril}
\bibfield{author}{\bibinfo{person}{Mubashir ul Islam}, \bibinfo{person}{Humza Sami}, \bibinfo{person}{Pierre-Emmanuel Gaillardon}, {and} \bibinfo{person}{Valerio Tenace}.} \bibinfo{year}{2024}\natexlab{}.
\newblock \bibinfo{title}{AIvril: AI-Driven RTL Generation With Verification In-The-Loop}.
\newblock
\newblock
\showeprint[arxiv]{2409.11411}~[cs.AI]
\urldef\tempurl%
\url{https://arxiv.org/abs/2409.11411}
\showURL{%
\tempurl}


\bibitem[Wan et~al\mbox{.}(2025)]%
        {genben}
\bibfield{author}{\bibinfo{person}{Gwok-Waa Wan}, \bibinfo{person}{Yubo Wang}, \bibinfo{person}{Sam-Zaak Wong}, \bibinfo{person}{Jia Xiong}, \bibinfo{person}{Qixiang Chen}, \bibinfo{person}{Jingyi Zhang}, \bibinfo{person}{Mingchi Zhang}, \bibinfo{person}{Tao Ni}, \bibinfo{person}{Mengnv Xing}, \bibinfo{person}{Yusheng Hua}, \bibinfo{person}{Zhe Jiang}, \bibinfo{person}{Nan Guan}, \bibinfo{person}{Ying Wang}, \bibinfo{person}{Ning Xu}, \bibinfo{person}{Qiang Xu}, {and} \bibinfo{person}{Xi Wang}.} \bibinfo{year}{2025}\natexlab{}.
\newblock \showarticletitle{{GenBen: A Generative Benchmark for LLM-Aided Design}}. In \bibinfo{booktitle}{\emph{{Arxiv}}}. \bibinfo{address}{Nanjing, China}.
\newblock
\urldef\tempurl%
\url{https://hal.science/hal-05098871}
\showURL{%
\tempurl}


\bibitem[Wang et~al\mbox{.}(2025a)]%
        {veridebug}
\bibfield{author}{\bibinfo{person}{Ning Wang}, \bibinfo{person}{Bingkun Yao}, \bibinfo{person}{Jie Zhou}, \bibinfo{person}{Yuchen Hu}, \bibinfo{person}{Xi Wang}, \bibinfo{person}{Nan Guan}, {and} \bibinfo{person}{Zhe Jiang}.} \bibinfo{year}{2025}\natexlab{a}.
\newblock \bibinfo{title}{VeriDebug: A Unified LLM for Verilog Debugging via Contrastive Embedding and Guided Correction}.
\newblock
\newblock
\showeprint[arxiv]{2504.19099}~[cs.SE]
\urldef\tempurl%
\url{https://arxiv.org/abs/2504.19099}
\showURL{%
\tempurl}


\bibitem[Wang et~al\mbox{.}(2025b)]%
        {veriseek}
\bibfield{author}{\bibinfo{person}{Ning Wang}, \bibinfo{person}{Bingkun Yao}, \bibinfo{person}{Jie Zhou}, \bibinfo{person}{Yuchen Hu}, \bibinfo{person}{Xi Wang}, \bibinfo{person}{Zhe Jiang}, {and} \bibinfo{person}{Nan Guan}.} \bibinfo{year}{2025}\natexlab{b}.
\newblock \showarticletitle{Large Language Model for Verilog Generation with Code-Structure-Guided Reinforcement Learning}. In \bibinfo{booktitle}{\emph{2025 IEEE International Conference on LLM-Aided Design (ICLAD)}}. \bibinfo{pages}{164--170}.
\newblock
\urldef\tempurl%
\url{https://doi.org/10.1109/ICLAD65226.2025.00025}
\showDOI{\tempurl}


\bibitem[Wei et~al\mbox{.}(2024)]%
        {OSS-Instruct}
\bibfield{author}{\bibinfo{person}{Yuxiang Wei}, \bibinfo{person}{Zhe Wang}, \bibinfo{person}{Jiawei Liu}, \bibinfo{person}{Yifeng Ding}, {and} \bibinfo{person}{Lingming Zhang}.} \bibinfo{year}{2024}\natexlab{}.
\newblock \showarticletitle{Magicoder: empowering code generation with OSS-INSTRUCT}. In \bibinfo{booktitle}{\emph{Proceedings of the 41st International Conference on Machine Learning}} (Vienna, Austria) \emph{(\bibinfo{series}{ICML'24})}. \bibinfo{publisher}{JMLR.org}, Article \bibinfo{articleno}{2158}, \bibinfo{numpages}{26}~pages.
\newblock


\bibitem[Xu et~al\mbox{.}(2025)]%
        {meic}
\bibfield{author}{\bibinfo{person}{Ke Xu}, \bibinfo{person}{Jialin Sun}, \bibinfo{person}{Yuchen Hu}, \bibinfo{person}{Xinwei Fang}, \bibinfo{person}{Weiwei Shan}, \bibinfo{person}{Xi Wang}, {and} \bibinfo{person}{Zhe Jiang}.} \bibinfo{year}{2025}\natexlab{}.
\newblock \showarticletitle{MEIC: Re-thinking RTL Debug Automation using LLMs}. In \bibinfo{booktitle}{\emph{Proceedings of the 43rd IEEE/ACM International Conference on Computer-Aided Design}}. \bibinfo{publisher}{Association for Computing Machinery}, \bibinfo{address}{New York, NY, USA}, Article \bibinfo{articleno}{100}, \bibinfo{numpages}{9}~pages.
\newblock
\showISBNx{9798400710773}
\urldef\tempurl%
\url{https://doi.org/10.1145/3676536.3676801}
\showURL{%
\tempurl}


\bibitem[Yang et~al\mbox{.}(2024)]%
        {strider}
\bibfield{author}{\bibinfo{person}{Deheng Yang}, \bibinfo{person}{Jiayu He}, \bibinfo{person}{Xiaoguang Mao}, \bibinfo{person}{Tun Li}, \bibinfo{person}{Yan Lei}, \bibinfo{person}{Xin Yi}, {and} \bibinfo{person}{Jiang Wu}.} \bibinfo{year}{2024}\natexlab{}.
\newblock \showarticletitle{Strider: Signal Value Transition-Guided Defect Repair for HDL Programming Assignments}.
\newblock \bibinfo{journal}{\emph{IEEE Transactions on Computer-Aided Design of Integrated Circuits and Systems}} \bibinfo{volume}{43}, \bibinfo{number}{5} (\bibinfo{year}{2024}), \bibinfo{pages}{1594--1607}.
\newblock
\urldef\tempurl%
\url{https://doi.org/10.1109/TCAD.2023.3341750}
\showDOI{\tempurl}


\bibitem[Yang et~al\mbox{.}(2025)]%
        {haven}
\bibfield{author}{\bibinfo{person}{Yiyao Yang}, \bibinfo{person}{Fu Teng}, \bibinfo{person}{Pengju Liu}, \bibinfo{person}{Mengnan Qi}, \bibinfo{person}{Chenyang Lv}, \bibinfo{person}{Ji Li}, \bibinfo{person}{Xuhong Zhang}, {and} \bibinfo{person}{Zhezhi He}.} \bibinfo{year}{2025}\natexlab{}.
\newblock \showarticletitle{HaVen: Hallucination-Mitigated LLM for Verilog Code Generation Aligned with HDL Engineers}. In \bibinfo{booktitle}{\emph{2025 Design, Automation \& Test in Europe Conference (DATE)}}. \bibinfo{pages}{1--7}.
\newblock
\urldef\tempurl%
\url{https://doi.org/10.23919/DATE64628.2025.10993072}
\showDOI{\tempurl}


\bibitem[Yao et~al\mbox{.}({[n.\,d.]})]%
        {ReAct}
\bibfield{author}{\bibinfo{person}{Shunyu Yao}, \bibinfo{person}{Jeffrey Zhao}, \bibinfo{person}{Dian Yu}, \bibinfo{person}{Nan Du}, \bibinfo{person}{Izhak Shafran}, \bibinfo{person}{Karthik Narasimhan}, {and} \bibinfo{person}{Yuan Cao}.} \bibinfo{year}{[n.\,d.]}\natexlab{}.
\newblock \showarticletitle{ReAct: Synergizing Reasoning and Acting in Language Models}.
\newblock \bibinfo{journal}{\emph{International Conference on Learning Representations (ICLR)}} (\bibinfo{year}{[n.\,d.]}).
\newblock
\urldef\tempurl%
\url{https://par.nsf.gov/biblio/10451467}
\showURL{%
\tempurl}


\bibitem[Yao et~al\mbox{.}(2025)]%
        {hdldebugger}
\bibfield{author}{\bibinfo{person}{Xufeng Yao}, \bibinfo{person}{haoyang Li}, \bibinfo{person}{Tsz~Ho Chan}, \bibinfo{person}{Wenyi Xiao}, \bibinfo{person}{Mingxuan Yuan}, \bibinfo{person}{Yu Huang}, \bibinfo{person}{Lei Chen}, {and} \bibinfo{person}{Bei Yu}.} \bibinfo{year}{2025}\natexlab{}.
\newblock \showarticletitle{HDLdebugger: Streamlining HDL debugging with Large Language Models}.
\newblock \bibinfo{journal}{\emph{ACM Trans. Des. Autom. Electron. Syst.}} (\bibinfo{date}{May} \bibinfo{year}{2025}).
\newblock
\showISSN{1084-4309}
\urldef\tempurl%
\url{https://doi.org/10.1145/3735638}
\showDOI{\tempurl}
\newblock
\shownote{Just Accepted}.


\bibitem[Ye et~al\mbox{.}(2025)]%
        {llmasjudge3}
\bibfield{author}{\bibinfo{person}{Ziyi Ye}, \bibinfo{person}{Xiangsheng Li}, \bibinfo{person}{Qiuchi Li}, \bibinfo{person}{Qingyao Ai}, \bibinfo{person}{Yujia Zhou}, \bibinfo{person}{Wei Shen}, \bibinfo{person}{Dong Yan}, {and} \bibinfo{person}{Yiqun LIU}.} \bibinfo{year}{2025}\natexlab{}.
\newblock \showarticletitle{Learning {LLM}-as-a-Judge for Preference Alignment}. In \bibinfo{booktitle}{\emph{The Thirteenth International Conference on Learning Representations}}.
\newblock
\urldef\tempurl%
\url{https://openreview.net/forum?id=HZVIQE1MsJ}
\showURL{%
\tempurl}


\bibitem[Zhang et~al\mbox{.}(2024)]%
        {llmsurvey2}
\bibfield{author}{\bibinfo{person}{Duzhen Zhang}, \bibinfo{person}{Yahan Yu}, \bibinfo{person}{Jiahua Dong}, \bibinfo{person}{Chenxing Li}, \bibinfo{person}{Dan Su}, \bibinfo{person}{Chenhui Chu}, {and} \bibinfo{person}{Dong Yu}.} \bibinfo{year}{2024}\natexlab{}.
\newblock \showarticletitle{{MM}-{LLM}s: Recent Advances in {M}ulti{M}odal Large Language Models}. In \bibinfo{booktitle}{\emph{Findings of the Association for Computational Linguistics: ACL 2024}}, \bibfield{editor}{\bibinfo{person}{Lun-Wei Ku}, \bibinfo{person}{Andre Martins}, {and} \bibinfo{person}{Vivek Srikumar}} (Eds.). \bibinfo{publisher}{Association for Computational Linguistics}, \bibinfo{address}{Bangkok, Thailand}, \bibinfo{pages}{12401--12430}.
\newblock
\urldef\tempurl%
\url{https://doi.org/10.18653/v1/2024.findings-acl.738}
\showDOI{\tempurl}


\bibitem[Zhang et~al\mbox{.}(2025)]%
        {hallucination}
\bibfield{author}{\bibinfo{person}{Ziyao Zhang}, \bibinfo{person}{Chong Wang}, \bibinfo{person}{Yanlin Wang}, \bibinfo{person}{Ensheng Shi}, \bibinfo{person}{Yuchi Ma}, \bibinfo{person}{Wanjun Zhong}, \bibinfo{person}{Jiachi Chen}, \bibinfo{person}{Mingzhi Mao}, {and} \bibinfo{person}{Zibin Zheng}.} \bibinfo{year}{2025}\natexlab{}.
\newblock \showarticletitle{LLM Hallucinations in Practical Code Generation: Phenomena, Mechanism, and Mitigation}.
\newblock \bibinfo{journal}{\emph{Proc. ACM Softw. Eng.}} \bibinfo{volume}{2}, \bibinfo{number}{ISSTA}, Article \bibinfo{articleno}{ISSTA022} (\bibinfo{date}{June} \bibinfo{year}{2025}), \bibinfo{numpages}{23}~pages.
\newblock
\urldef\tempurl%
\url{https://doi.org/10.1145/3728894}
\showDOI{\tempurl}


\bibitem[Zhu et~al\mbox{.}(2025)]%
        {llmasjudge2}
\bibfield{author}{\bibinfo{person}{Lianghui Zhu}, \bibinfo{person}{Xinggang Wang}, {and} \bibinfo{person}{Xinlong Wang}.} \bibinfo{year}{2025}\natexlab{}.
\newblock \showarticletitle{Judge{LM}: Fine-tuned Large Language Models are Scalable Judges}. In \bibinfo{booktitle}{\emph{The Thirteenth International Conference on Learning Representations}}.
\newblock
\urldef\tempurl%
\url{https://openreview.net/forum?id=xsELpEPn4A}
\showURL{%
\tempurl}


\end{thebibliography}

\end{document}